\documentclass[oupdraft]{bio}
\usepackage{url}
\usepackage{bm}
\usepackage{booktabs}
\usepackage{threeparttable}
\usepackage{algorithm, algpseudocode}
\usepackage{setspace}

\newcommand{\bbeta}{{\bm \beta}}

\newcommand{\bnu}{\bm{\nu}}

\newcommand{\bSigma}{\bm{\Sigma}}

\newcommand{\bOmega}{\mathbf{\Omega}}

\newcommand{\bD}{\mathbf{D}}

\newcommand{\bW}{\mathbf{W}}
\newcommand{\bX}{\mathbf{X}}

\newcommand{\bZ}{\mathbf{Z}}

\newcommand{\bz}{\mathbf{z}}

\newcommand{\cD}{\mathcal{D}}

\newcommand{\cM}{\mathcal{M}}

\newcommand{\cT}{\mathcal{T}}

\newcommand{\biy}{\boldsymbol{y}}

\newcommand{\bmu}{\bm{\mu}}

\newcommand{\bzero}{\bm{0}}

\begin{document}

\title{Modeling Joint Health Effects of Environmental Exposure Mixtures with Bayesian Additive Regression Trees}

\author{JACOB R. ENGLERT$^\ast$ \\
\textit{Department of Biostatistics and Bioinformatics, Emory University, 1518 Clifton Rd NE, Atlanta, GA 30322, USA} \\
jacob.englert@emory.edu \\
STEFANIE T. EBELT \\ 
\textit{Gangarosa Department of Environmental Health, Emory University, 1518 Clifton Rd NE, Atlanta, GA 30322, USA}\\
HOWARD H. CHANG \\
\textit{Department of Biostatistics and Bioinformatics, Emory University, 1518 Clifton Rd NE, Atlanta, GA 30322, USA}
}

\markboth%
{J. R. Englert and others}
{Modeling Joint Health Effects of Environmental Exposure Mixtures with BART}

\maketitle

\footnotetext{To whom correspondence should be addressed.}

\begin{abstract}
{Studying the association between mixtures of environmental exposures and health outcomes can be challenging due to issues such as correlation among the exposures and non-linearities or interactions in the exposure-response function. For this reason, one common strategy is to fit flexible nonparametric models to capture the true exposure-response surface. However, once such a model is fit, further decisions are required when it comes to summarizing the marginal and joint effects of the mixture on the outcome. In this work, we describe the use of soft Bayesian additive regression trees (BART) to estimate the exposure-risk surface describing the effect of mixtures of chemical air pollutants and temperature on asthma-related emergency department (ED) visits during the warm season in Atlanta, Georgia from 2011-2018. BART is chosen for its ability to handle large datasets and for its flexibility to be incorporated as a single component of a larger model. We then summarize the results using a strategy known as accumulated local effects to extract meaningful insights into the mixture effects on asthma-related morbidity. Notably, we observe negative associations between NO$_2$ and asthma ED visits and harmful associations between ozone and asthma ED visits, both of which are particularly strong on lower temperature days.}
{Accumulated local effects; Air pollution; Asthma; Bayesian additive regression trees; Environmental mixtures}
\end{abstract}

\section{Introduction}
\label{sec:intro}

Asthma affected an estimated 25 million (7.7\%) individuals in the United States in the year 2021 per the National Health Interview Survey 
\citep{centers_for_disease_control_and_prevention_most_2021}. In that same year, the Healthcare Cost and Utilization Project estimated a total of 5.8 million asthma-related emergency department (ED) visits, of which 1.4 million required hospitalization and 930,000 listed asthma as the primary diagnosis \citep{agency_for_healthcare_research_and_quality_rockville_md_hcupnet_2021}. In this work we are interested in studying the marginal and joint associations between elevated concentrations of multiple airborne chemical pollutants on asthma-related ED visit rates in Atlanta, Georgia.

Previous studies of Atlanta and other U.S. cities have found harmful associations between asthma-related ED visits and environmental pollutants such as fine particulate matter of equal to or less than 2.5~µm in diameter (PM$_{2.5}$), nitrogen dioxide (NO$_2$), ozone (O$_3$), and carbon monoxide (CO), among others \citep{strickland_short-term_2010, ji_meta-analysis_2011, winquist_joint_2014, alhanti_ambient_2016, strickland_pediatric_2016, olenick_assessment_2017, bi_acute_2023}. The majority of these studies analyze exposures individually due to the challenges associated with modeling mixtures of correlated exposures. However, studies do often stratify analyses to study effect modification. For example, \cite{olenick_assessment_2017} reported that neighborhood-level socioeconomic status may affect the association between between air pollution and pediatric asthma morbidity. Additionally, some studies have also reported modification of the effect of ozone on mortality by temperature \citep{wilson_modeling_2014}. 

Modeling of environmental mixtures is often framed in one of two ways: targeting a restricted class of research questions using easily interpretable parametric models, or estimating the true exposure-response surface with fancier (but less interpretable) models based on Gaussian processes, regression tree ensembles, etc. The former includes summary index approaches such as quantile G-computation \citep{keil_quantile-based_2020} and weighted quantile sum \citep{carrico_characterization_2015}, while the latter encompasses tools like Bayesian kernel machine regression (BKMR) \citep{bobb_bayesian_2015}, and more recently treed distributed lag mixture models (TDLMM) \citep{mork_estimating_2023} and multiple exposure distributed lag models \citep{antonelli_multiple_2023}. BKMR is most useful for estimating smooth exposure-response functions containing interactions and non-linearities, while the latter two focus primarily on interaction and lagged effects over discrete time intervals. In our review, existing methods typically address at most two of 1) non-linearity, 2) interaction, or 3) lagged effects (see \cite{wilson_kernel_2022} for an approach which seeks to address all three). We propose using soft Bayesian additive regression trees (BART) \citep{linero_bayesian_2018-1} as an alternative to the BKMR approach for estimating interactions and non-linearities. While soft BART is computationally slower than traditional BART, the tree-based approach is more feasible than BKMR when working with datasets with a large number of observations, as in the motivating Atlanta dataset.

Following the fit of a mixture model, summarization of the estimated exposure-risk function with respect to one or two exposures often involves fixing the other exposures within the mixture to some quantile, or uses partial dependence statistics \citep{friedman_greedy_2001}. These approaches tend to extrapolate to implausible mixture levels in the estimation process, particularly in the context of correlated exposures. We propose using accumulated local effects \citep{apley_visualizing_2020} as an alternative approach that avoids this issue and has other benefits as well.

The main contribution of this work is to leverage a modeling approach based on Bayesian regression tree ensembles and subsequent summarization strategy for evaluating the effects of multi-pollutant mixtures on asthma morbidity in the city of Atlanta. We introduce the data for this application in Section \ref{sec:data} and outline the methodology in Section \ref{sec:methods}. We then demonstrate the utility of the proposed approach though a simulation study (Section \ref{sec:simulation}) before finally presenting our main findings from the application in Section \ref{sec:application}.

\section{Data}
\label{sec:data}

\subsection{Health Data}
Patient-level billing records for ED visits to hospitals in the metropolitan Atlanta area from 2011-2018 were obtained from the Georgia Hospital Association. These data included admission date, billing address, International Classification of Disease (ICD) version 9 or 10 discharge diagnosis codes, and various patient characteristics. We restricted the ED visit data to only include visits containing an asthma diagnosis (ICD-9 code 493; ICD-10 code J45) and occurring during Atlanta's warm season (April-October). These visits were then aggregated by ZIP code and date for the analysis. These data have previously been used for various asthma and air pollution association studies \citep{strickland_short-term_2010, strickland_pediatric_2016, olenick_assessment_2017, lappe_pollen_2023, bi_acute_2023}.

\subsection{Air Pollution Data}
Air pollution concentration data are collected daily and include fine particulate matter with diameter 2.5 µm and smaller (PM$_{2.5}$, 24-hr average, µm/m$^3$), ozone (O$_3$, 8-hr max, ppm), nitrogen dioxide (NO$_2$, 1-hr max, ppb), and carbon monoxide (CO, 1-hr max, ppb). The estimates are derived from the data fusion model described by \cite{senthilkumar_using_2022}, which utilized simulations from the Community Multiscale Air Quality Model monitoring data from the Environmental Protection Agency's Air Quality System database. The data product is available at a 12km gridded spatial resolution and is linked to each ZIP code based on area-weighted averaging.

\subsection{Other Data}
Maximum daily temperature is obtained from Daymet \citep{thornton_daymet_2022}. The 1km gridded product was spatially averaged within each ZIP code, and linked to the ED visit data by both date and ZIP code. Annual ZIP code-level estimates of total population and the percent of the population below the poverty level are obtained from the 5-year American Community Survey for years 2011-2018.

\section{Methods}
\label{sec:methods}

\subsection{Soft BART}
Bayesian additive regression trees (BART) is a nonparametric machine learning approach which approximates complicated functions using sums of shallow Bayesian decision trees \citep{chipman_bart_2010}. In this sense, the approach is similar to \emph{boosting} from the machine learning literature. In recent years BART has soared in popularity due to its performance on prediction, classification, and causal inference tasks \citep{hill_bayesian_2020}. Ultimately, BART is a tree-based approach that can only approximate smooth functions with rigid fits. \cite{linero_bayesian_2018-1} propose a \emph{soft} version of BART, which adapts to smooth functions better than traditional BART by swapping traditional decision trees for soft decision trees \citep{irsoy_soft_2012}. In a soft decision tree, the prediction for an observation is a weighted average of all of the leaf node parameters, where the weights are defined as the probability an observation is mapped to each leaf node as determined by, say, a logistic gating function \citep{linero_bayesian_2018-1}. When compared to the deterministic predictions obtained from a traditional decision tree, this has the effect of smoothing over the otherwise rigid decision rules that form the binary tree. Since we generally expect the exposure-risk surface to be smooth, we opt to use this version of BART in our implementation.

\subsection{Negative Binomial Regression with BART}
We propose using the following Poisson regression model with gamma-distributed mean:
\begin{gather}
    Y_{ij} \mid \theta_{ij} \sim \text{Poisson}(\theta_{ij}) \nonumber \\
    \theta_{ij} \sim \text{Gamma}(\xi, \exp{(\eta_{ij})}) \nonumber \\
    \eta_{ij} = \log(\text{Pop}_{ij}) + \bX_{ij}^T \bbeta + f(\bZ_{ij}) + \nu_i, \label{eq:main-model}
\end{gather}
where $Y_{ij}$ and $\theta_{ij}$ represent the observed and expected asthma ED counts in region $i$ on day $j$, respectively, for $i = 1, \ldots, I$ and $j = 1, \ldots, J$. It follows that marginally, $Y_{ij}$ is distributed as a negative binomial random variable with parameters $\xi$ and $p_{ij} = \exp{(\eta_{ij})} / \left[1 + \exp{(\eta_{ij})} \right]$. The log expected counts are offset by $\text{Pop}_{ij}$, the population of region $i$ at time $j$, and the overdispersion in the counts is represented by $\xi$. Potential confounders such as federal holidays and socioeconomic factors are represented by the $p \times 1$ vector $\bX_{ij}$, while all exposures are represented by $q \times 1$ vector $\bZ_{ij}$. To account for additional unexplained variation in the counts due to location we include a ZIP code specific random intercept $\nu_i$.

The confounders are modeled linearly with regression coefficients given by the $p \times 1$ vector $\bbeta$. The exposures are modeled using soft BART \citep{linero_bayesian_2018-1}. Specifically, in Equation \eqref{eq:main-model} we use $f(\bZ_{ij}) = \sum_{t=1}^T g(\bZ_{ij}; \cT_t, \cM_t)$. The $\left\{\cT_t, \cM_t\right\}_{t=1}^T$ parameters correspond to the tree structures and scalar-valued leaf node parameters associated with the soft BART model, and $g$ is the function which maps a set of exposures $\bZ_{ij}$ to its prediction from a single soft decision tree. Our approach is similar to that used in \citet{mutiso_bayesian_2024}, with the main difference being that we substitute the BKMR, which uses Gaussian processes, for soft BART.

\subsection{Model Estimation}
The model described in Equation \eqref{eq:main-model} is estimated with a Markov chain Monte-Carlo (MCMC) algorithm consisting of a mix of Gibbs and Metropolis-Hastings (MH) steps. BART (and by extension, soft BART), has been extended to many different outcome types since the original BART model was proposed. However, these extensions typically require conditional conjugacy between the outcome distribution and prior distribution on the individual leaf node parameters to facilitate the sampling of tree structures. Since the negative binomial likelihood does not itself admit a conditionally conjugate prior, we adopt the framework proposed by \citet{pillow_fully_2012}. We augment the observed outcome $y_{ij}$ with latent weights $\omega_{ij}$ sampled from a Pólya-gamma (PG) distribution \citep{polson_bayesian_2013}. Specifically, if we independently draw $\omega_{ij} \sim \text{PG}(y_{ij} + \xi, \eta_{ij})$, then the latent outcome $y^*_{ij} = \frac{y_{ij} - \xi}{2\omega_{ij}}$ follows a normal distribution with mean $\eta_{ij}$ and variance $1 / \omega_{ij}$. This leads to a convenient Gibbs sampler based on the Bayesian backfitting approach of \citet{hastie_bayesian_2000} for updating $\bbeta$, $\left\{\cM_t \right\}_{t=1}^T$, and $\bnu$, when multivariate normal prior distributions are chosen. The full forms of these conjugate updates are provided in Appendix A of the Supplementary Material.

In BART, new tree structures are sampled from their marginal distribution and updated using an MH step by first integrating out the leaf node parameters. The tree structures themselves are proposed from a so-called \emph{branching process} prior, which grow, prune, or perturb the existing structures from the previous iteration \citep{chipman_bart_2010, pratola_efficient_2016}. A detailed explanation of how this approach works with weights, such as those introduced by the Pólya-gamma data augmentation scheme, is outlined in \citet{bleich_bayesian_2014}. The tree prior used for soft BART is more involved, optionally including hyperparameters and hyperpriors responsible for the degree of smoothness and/or sparsity, but the general concept is the same. One of the benefits of BART is that default priors tend to work well in a variety of circumstances. We use default priors for updating $\left\{\cT_t, \cM_t \right\}_{t=1}^T$ as detailed in \citet{linero_bayesian_2018-1}.

We assign a mean-zero proper conditional autoregressive (pCAR) prior for the ZIP code-level random intercepts $\bnu = (\nu_1, \ldots, \nu_I)^T$. In mathematical terms, $\bnu \sim \text{MVN}\left(\bzero, \bSigma_\nu \right)$ with $\bSigma_\nu = \tau^2(\bD - \rho \bW)^{-1}$, where $\bW$ is the $I \times I$ first-order adjacency matrix, and $\bD$ is the $I \times I$ diagonal matrix containing the number of neighbors for each region. This allows for a portion of the variability in the response unexplained by the predictors to be attributed to unmeasured spatial factors. We assign an inverse Gamma and discrete uniform hyperprior to $\tau^2$ and $\rho$, respectively.

Lastly, we update the dispersion parameter $\xi$ using the conjugate sampling routine described in \cite{zhou_lognormal_2012}. This technique relies on expressing the $Y_{ij} \sim \text{NB}(\xi, p_{ij})$ marginal distribution as a compound Poisson distribution. The details for this step and the others described in this section are outlined in Appendix A of the Supplementary Material. A summary of a single MCMC iteration is provided in Algorithm \ref{alg:mcmc}.

\begin{algorithm}[!h]
\caption{One MCMC iteration of the Soft BART negative binomial model}
\label{alg:mcmc}
\begin{algorithmic}[1]
\State \textbf{Input:} $\cD = \left\{\biy, \bX, \bZ, \text{Population}, \bD, \bW \right\}, \bbeta, \bnu, \tau^2, \rho, \left\{\cT_t, \cM_t \right\}_{t=1}^T, \xi, \alpha_\tau, \beta_\tau, \alpha_\xi, \beta_\xi$
\For{$i = 1, \ldots, I$ and $j = 1, \ldots, J$}
    \State Draw $\omega_{ij} \sim \text{PG}(y_{ij} + \xi, \eta_{ij})$.
    \State Compute $y^*_{ij} = \frac{y_{ij} - \xi}{2\omega_{ij}}$.
\EndFor
\State Form $\bOmega = \text{diag}(\omega_{11}, \ldots, \omega_{1,J}, \ldots, \omega_{I1}, \ldots, \omega_{IJ})$.
\State Draw $\bbeta \sim \text{MVN} \left(\bmu^*_\beta, \bSigma^*_\beta \right)$, where $\bmu^*_\beta$ and $\bSigma^*_\beta$ are defined as in Appendix A.3.
\State Draw $\bnu \sim \text{MVN} \left(\bmu^*_\nu, \bSigma^*_\nu \right)$, where $\bmu^*_\nu$ and $\bSigma^*_\nu$ are defined as in Appendix A.4.
\State Draw $\tau^2 \sim \text{IG} \left(\alpha_\tau + I / 2, \beta_\tau + \bnu^T(\bD - \rho \bW) \bnu) / 2 \right)$
\State Draw $\rho$ from its discrete posterior distribution described in Appendix A.6.
\For{$t = 1, \ldots, T$}
    \State \parbox[t]{\dimexpr\linewidth-\algorithmicindent}{Propose/update $\cT_t, \cM_t$, and any associated hyperparameters governing the degree of smoothness and/or sparsity as described in \cite{linero_bayesian_2018-1}.}
\EndFor
\State Draw $L_{ij} \sim \text{CRT}(\xi, y_{ij})$ for $i = 1, \ldots, I$, $j = 1, \ldots, J$.
\State Draw $\xi \sim \text{Gamma}\left(\alpha_\xi + \sum_{i=1}^I \sum_{j=1}^J L_{ij}, \beta_\xi - \sum_{i=1}^I \sum_{j=1}^J \ln(1-p_{ij}) \right)$.
\end{algorithmic}
\end{algorithm}

\subsection{Model Interpretation with Accumulated Local Effects}
When using flexible methods which target the response-surface directly, interpretation of the resulting fit requires just as much thought as the estimation itself. In the environmental mixtures setting, it is common to focus on the marginal effect of a single exposure on the outcome by evaluating the exposure-response function at several levels of the chosen exposure and plotting the result. This strategy can also be used for studying the joint effects of two exposures using, say, contour plots. Analyzing/visualizing joint effects of more than two continuous exposures is rather difficult, and thus is not as common. Mathematically, for an exposure-response function in which we are interested in evaluating the effect of $\bZ_k$ on $f$, we evaluate $\hat{f}\left(\bz_k \mid \bZ_{[-k]}\right)$ for several values $\bz_k$ in the observed range of $\bZ_k$.

In settings where there are more than two exposures, a decision must be made regarding the treatment of the ``other" exposures $\bZ_{[-k]}$ when evaluating the exposure-response function for 1-2 exposures of interest $\bZ_k$. A common choice is to set $\bZ_{[-k]}$ to some fixed value (e.g., their observed medians) while varying $\bZ_k$. In fact, one might set $\bZ_{[-k]}$ to multiple values (e.g., median and 95th percentile), and plot $\hat{f}\left(\bz_k \mid \bZ_{[-k]}\right)$ for each setting. This approach is not ideal since the exposure-response function ultimately depends on the selected values for $\bZ_{[-k]}$, of which there are many choices for each exposure in the model - none of which are the perfect choice, and many of which are poor choices.

Another option is to calculate partial dependence (PD) functions \citep{friedman_greedy_2001}. PD functions target the average exposure effect across the marginal distribution of the observed data, setting $\bZ_{[-k]}$ to their observed values $\bZ_{ij, [-k]}$. In this manner, PD functions avoid having to make a choice regarding the values of $\bZ_{[-k]}$, and the resulting estimates naturally incorporate the variability in $\bZ_{[-k]}$ across specified values of $\bZ_k$. One of the major limitations of PD functions is that they are computationally burdensome, requiring predictions for every observation in the study at every value considered for $\bZ_k$.

Both the fixed-value and PD approaches run the risk of extrapolating when evaluating $\hat{f}\left(\bz_k \mid \bZ_{[-k]}\right)$, particularly if the exposures are correlated (which they often are). By this we mean that some of the evaluations of $\hat{f}\left(\bz_k \mid \bZ_{[-k]}\right)$ are made on implausible exposure profiles. This is a general issue for assessing covariate effects in black-box supervised learning models. One approach that has been proposed to combat this issue is accumulated local effects (ALE, \cite{apley_visualizing_2020}). The estimands for the partial effect of a single exposure $Z_k$ at some level $z_k$ for each of the three approaches are provided in Equations \eqref{eq:fixed-estimand}, \eqref{eq:pd-estimand}, and \eqref{eq:ale-estimand}.
\begin{gather}
    f_{k,Fixed}(z_k) \equiv f(z_k, \bz_{[-k]}) \label{eq:fixed-estimand} \\
    f_{k,PD}(z_k) \equiv \mathbb{E} \left[\ f(z_k,\bZ_{[-k]}) \right] \label{eq:pd-estimand} \\
    f_{k,ALE}(z_k) \equiv \int_{z_{min,k}}^{z_k} \mathbb{E} \left[\frac{\partial f}{\partial Z_k}(Z_k,\bZ_{[-k]}) \mid Z_k = z'_k \right] d z'_k \label{eq:ale-estimand}
\end{gather}

For estimation purposes, Equation \eqref{eq:fixed-estimand} only requires on evaluation of $\hat{f}$ at some fixed value $\left(z_k, \bz_{[-k]}\right)$, Equation \eqref{eq:pd-estimand} requires an evaluation of $\hat{f}$ at $\left(z_k, \bZ_{ij,[-k]}\right)$ for all $i,j$, and Equation \eqref{eq:ale-estimand} requires an evaluation of $\hat{f}$ at $\left(z_k^{lower}, \bZ_{ij,[-k]}\right)$ and $\left(z_k^{upper}, \bZ_{ij,[-k]}\right)$ for all $i,j$ s.t. $Z_{ij,k} \in \left(z_k^{lower}, z_k^{upper}\right)$. The latter is a result of approximating $\frac{\partial f}{\partial Z_k}$ with small finite differences. For more information regarding ALE computation, we refer the reader to \cite{apley_visualizing_2020}. When using a Bayesian approach, each of Equations \eqref{eq:fixed-estimand}, \eqref{eq:pd-estimand}, and \eqref{eq:ale-estimand} would be evaluated for each sample from the posterior distribution. This allows one to obtain point-wise posterior means and uncertainty estimates at each $z_k$.

The benefits of using ALE over the other approaches are threefold: 1) effects of correlated exposures are isolated by targeting the partial derivative of $f$, 2) estimates are only informed by predictions on plausible exposure profiles since averaging is done using the conditional distribution (i.e., no extrapolation), and 3) the computation is relatively fast compared to PD functions since only two predictions are needed for each observation, regardless of the number of levels of $Z_k$ being considered.

\section{Simulation Study}
\label{sec:simulation}

To evaluate the proposed approach, specifically to different BART specifications, we conduct a brief simulation study. We use the populations and locations of the 128 ZIP codes from the first year (2011) of the application. We set confounder effects $\bbeta = (-2, -1, 1, 2)^T$ and the true exposure-risk surface function to $f(\bz) = -10 + \frac{f_0(\bz)}{5}$, where $f_0(\bz) = 10\sin(z_1z_2) + 20(z_3 - 0.5)^2 + 10z_4 + 5z_5$ is the benchmark function proposed in \cite{friedman_multivariate_1991}. While this surface only depends on five exposures, we generate five additional noise exposures (ten total exposures) to assess the performance in settings where not all exposures are important. Spatial random effects are sampled from their pCAR prior with $\rho = 0.9$ and $\tau^2 = 0.3$. For $i = 1, \ldots, 128$ regions and $j = 1, \ldots, 300$ observations, we simulate outcomes using the following data generating process:
\begin{enumerate}
    \item Generate $X_{ij,1}, X_{ij,2}, X_{ij,3}, X_{ij,4} \overset{i.i.d.}{\sim} \text{Uniform}(0, 1)$.
    \item Generate $\bZ_{ij} \sim \text{MVN}(\bzero_{10 \times 1}, \bSigma_{10 \times 10})$, where $\bSigma_{[1:5,1:5]}$ matches the observed correlation matrix of PM$_{2.5}$, NO$_2$, O$_3$, CO, and maximum temperature from the application. Only the first five exposures are used to generate the outcome. All values are scaled to $[0, 1]$ using min-max normalization.
    \item Sample $Y_{ij} \sim \text{NB} \left( \xi = 1, p_{ij} = \frac{e^{\eta_{ij}}}{1 + e^{\eta_{ij}}} \right)$ where $\eta_{ij} = \log\left(\text{Pop}_{ij} \right) + \bX_{ij}^T \bbeta + f(\bZ_{ij}) + \nu_i$
\end{enumerate}

We consider ensembles of size $T = \left\{10, 25, 50, 100\right\}$, both hard and soft decision rules, and both the classic and sparse branching processes \citep{linero_bayesian_2018}. Each setting is repeated 200 times. The average bias, root mean squared error (RMSE), and 95\% credible interval coverage for $f(\bz)$ are presented in Table \ref{tab:sim-bart-stats}, along with Monte Carlo standard error estimates.

\begin{table}

\caption{\label{tab:sim-bart-stats}Average bias, average 95\% CrI coverage,
      and RMSE for BART predictions across simulations.}
\centering
\begin{tabular}[t]{cccccc}
\toprule
$T$\footnotemark[1] & Soft\footnotemark[2] & Sparse\footnotemark[3] & Bias $\times 10$ (MCSE) & Coverage (MCSE) & RMSE $\times 10$ (MCSE)\\
\midrule
10 &  &  & 0.11 (0.0138) & 0.88 (0.0229) & 1.17 (0.0050)\\
10 &  & \checkmark & 0.11 (0.0142) & 0.96 (0.0143) & 1.06 (0.0043)\\
10 & \checkmark &  & 0.02 (0.0139) & 0.91 (0.0207) & 0.41 (0.0065)\\
10 & \checkmark & \checkmark & 0.02 (0.0137) & 0.92 (0.0188) & 0.41 (0.0061)\\
\addlinespace
25 &  &  & 0.05 (0.0141) & 0.78 (0.0293) & 0.98 (0.0037)\\
25 &  & \checkmark & 0.05 (0.0138) & 0.89 (0.0221) & 0.89 (0.0039)\\
25 & \checkmark &  & 0.01 (0.0139) & 0.94 (0.0167) & 0.39 (0.0056)\\
25 & \checkmark & \checkmark & 0.01 (0.0137) & 0.95 (0.0155) & 0.37 (0.0056)\\
\addlinespace
50 &  &  & 0.02 (0.0136) & 0.81 (0.0279) & 0.89 (0.0035)\\
50 &  & \checkmark & 0.02 (0.0139) & 0.88 (0.0228) & 0.82 (0.0037)\\
50 & \checkmark &  & 0.01 (0.0137) & 0.95 (0.0147) & 0.42 (0.0050)\\
50 & \checkmark & \checkmark & 0.02 (0.0140) & 0.96 (0.0140) & 0.37 (0.0054)\\
\addlinespace
100 &  &  & 0.01 (0.0140) & 0.88 (0.0233) & 0.85 (0.0035)\\
100 &  & \checkmark & 0.01 (0.0137) & 0.92 (0.0188) & 0.77 (0.0037)\\
100 & \checkmark &  & 0.02 (0.0138) & 0.96 (0.0131) & 0.46 (0.0047)\\
100 & \checkmark & \checkmark & 0.01 (0.0136) & 0.96 (0.0130) & 0.39 (0.0054)\\
\bottomrule
\multicolumn{6}{l}{\rule{0pt}{1em}CrI: Bayesian posterior credible interval.}\\
\multicolumn{6}{l}{\rule{0pt}{1em}MCSE: Monte Carlo Standard Error.}\\
\multicolumn{6}{l}{\rule{0pt}{1em}\textsuperscript{*} Number of trees.}\\
\multicolumn{6}{l}{\rule{0pt}{1em}\textsuperscript{\dag} Soft BART used \citep{linero_bayesian_2018-1}.}\\
\multicolumn{6}{l}{\rule{0pt}{1em}\textsuperscript{\ddag} Sparse branching process used \citep{linero_bayesian_2018}.}\\
\end{tabular}
\end{table}

Soft BART had excellent bias, coverage, and RMSE even when few trees were used. When many trees were used ($T = 100$), performance of traditional BART improved, but was still worse than soft BART in terms of coverage and RMSE. In general, increasing the number of trees beyond 25 did not appear to improve the performance of soft BART. Additionally, using the sparse branching process prior mostly resulted in improved coverage and reduced RMSE (Table \ref{tab:sim-bart-stats}). This suggests the sparsity-inducing Dirichlet prior was effective at identifying the important exposures and avoiding tree splitting rules based on the noise exposures.

\begin{figure}
    \centering
    \includegraphics[width=0.9\linewidth]{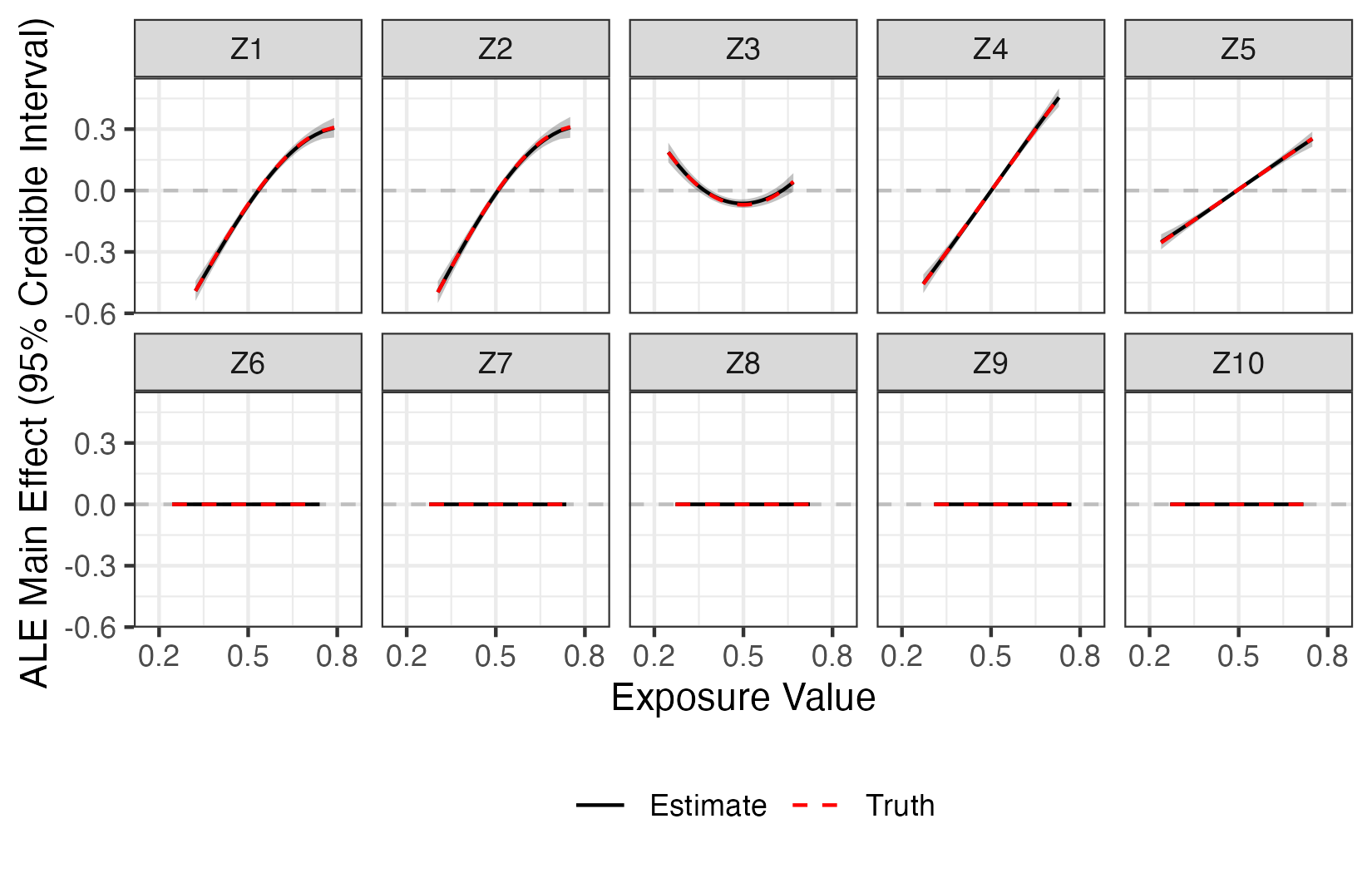}
    \caption{ALE main effects from the simulation study with $T=25$ soft, sparse trees. The solid black line represents the average ALE point estimate, and the gray ribbon represents the average 95\% posterior credible interval bounds across 200 simulations. ALEs are calculated using $K=40$ quantile intervals for each exposure. Plots are trimmed so that only the central 95\% of each exposure is displayed.}
    \label{fig:sim-study-main-effects}
\end{figure}

We present simulation results for the marginal ALE plots for each exposure using the $T = 25$ soft, sparse trees setting in Figure \ref{fig:sim-study-main-effects}. On average across simulations, the true functional forms of the five important exposures is recovered remarkably well. The null effects of the noise exposures are also accurately captured, primarily due to the ensembles avoiding splitting on these covariates entirely. Similar results for the pairwise ALE plots are included in Appendix B of the Supplementary Material.

In addition to the excellent performance on the recovery of $f$, estimates for $\bbeta$, $\xi$, $\tau^2$, $\rho$, and $\bnu$ were also generally unbiased and exhibited reasonable 95\% credible interval coverage. Results for these parameters are included in Appendix B of the Supplementary Material as well.

\section{Application}
\label{sec:application}

We observed 478,311 asthma-related ED visits from 219,136 daily counts during the warm season in Atlanta from 2011-2018. These visits came from 128 ZIP codes from Clayton, DeKalb, Gwinnett, Fulton, and Cobb counties. The number of asthma-related ED visits was relatively stable year-over-year during this time frame, but in general more visits are observed at either end of the warm season (April and October), and occasionally coincide with federal holidays as well. To account for potential confounding by these, we included an indicator variable representing federal holidays and a natural cubic spline on the day-of-year with 7 degrees of freedom per year (one per each warm season month). Given the previous findings of \citet{olenick_assessment_2017} suggesting the importance of socioeconomic status in this same dataset, we also include the annual ZIP code-level percent below the poverty threshold as a time-varying linear confounder.

We consider four chemical exposures PM$_{2.5}$, NO$_2$, O$_3$, and CO, as well as a meteorological exposure in maximum temperature. Each of these are recorded daily and included as 3-day moving averages. We fit soft BART ensembles of 10, 25, 50, and 100 trees for each of the five primary exposures individually and as a mixture. We run each model for 5,000 burn-in iterations, and then draw 1,000 posterior samples using a thinning interval of 10 iterations (15,000 total MCMC iterations).

\begin{figure}
    \centering
    \includegraphics[width=0.8\linewidth]{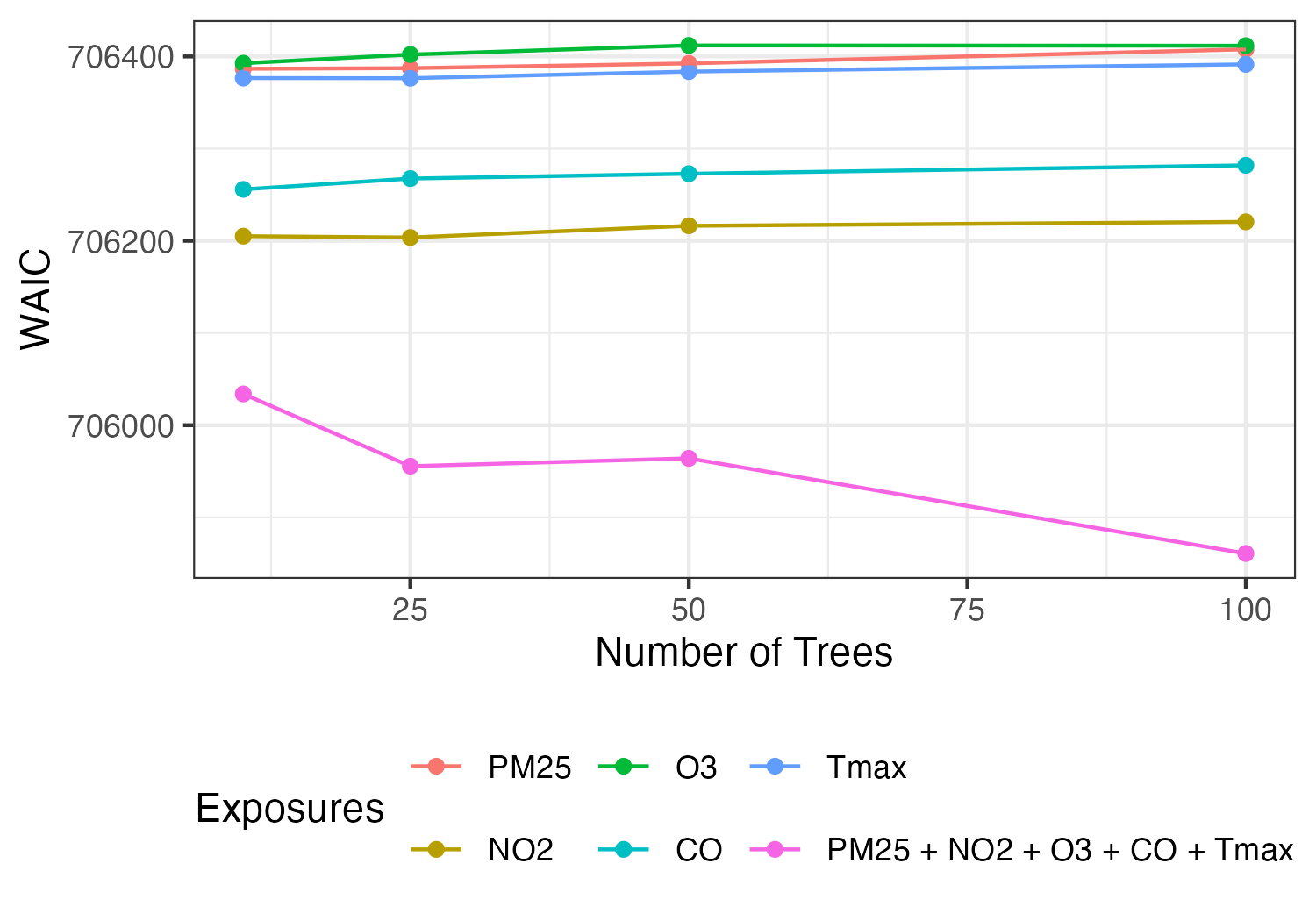}
    \caption{Widley applicable information criterion (WAIC) \citep{watanabe_asymptotic_2010, gelman_understanding_2014} for the single-exposure and mixture models fit to the Atlanta ED visit data.}
    \label{fig:waic-asthma}
\end{figure}

For each model, we compute the widely applicable information criterion (WAIC) as an approximation to leave-one-out cross-validation \citep{watanabe_asymptotic_2010, gelman_understanding_2014}. The results are plotted in Figure \ref{fig:waic-asthma}. For the single-exposure models, the WAIC is lowest for NO$_2$ and CO, suggesting that these two exposures are the most predictive of asthma-related ED visits when considered individually. As suspected, the mixture model containing all five exposures had a much lower WAIC than any of the single-exposure models. While increasing the ensemble size beyond 25 trees does not appear to improve the WAIC for any of the single-exposure models, larger ensembles may lead to some improved performance of the mixture model in terms of WAIC. However, when summarizing results of the mixture models with larger ensembles, we found the main findings to be generally similar to the fit with 25 trees. Due to this finding and for the sake of an even comparison, we will consider only the 25-tree single-exposure and mixture models in this section.

\begin{figure}
    \centering
    \includegraphics[width=0.6\linewidth]{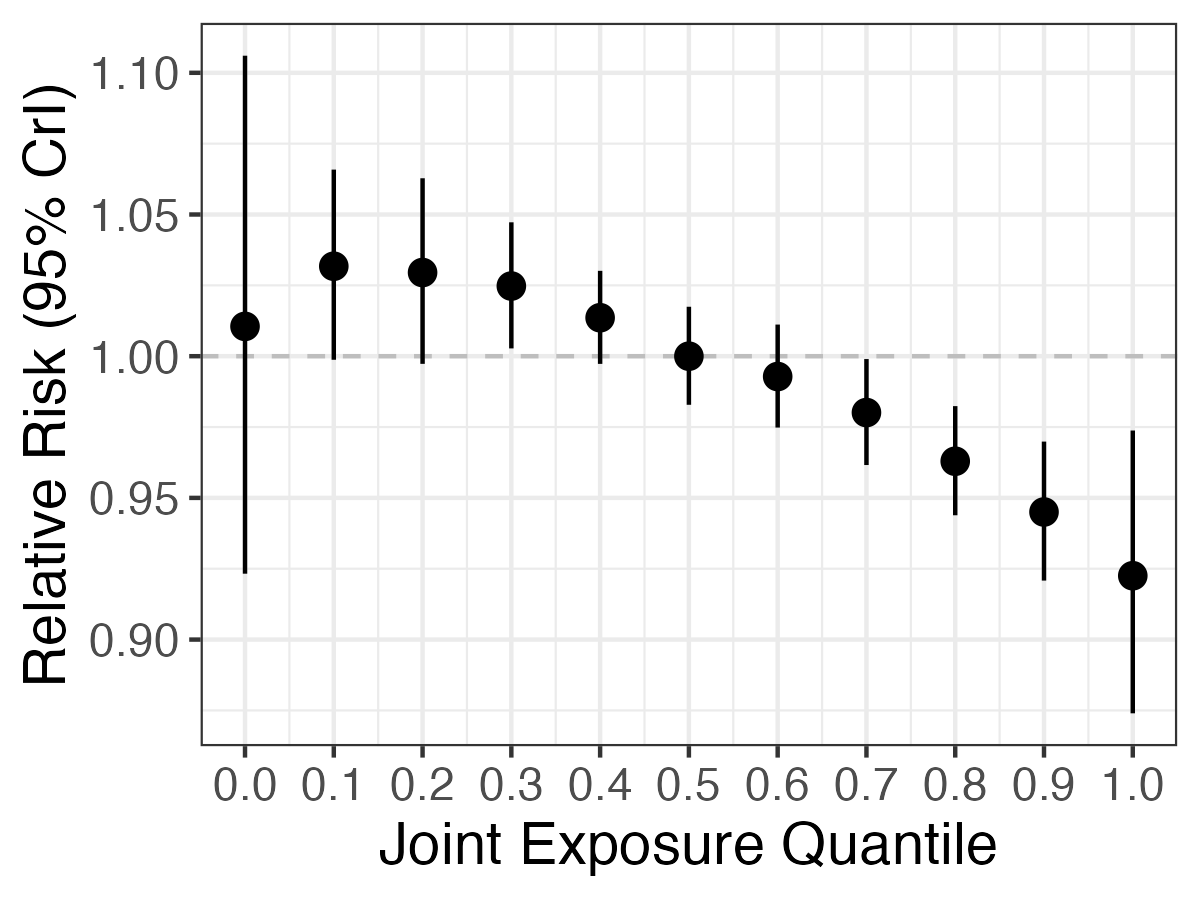}
    \caption{Estimated relative risk (and 95\% posterior credible interval) when all exposures are simultaneously set to the same decile for range of deciles. Estimate is relative to the risk when all exposures are set to their observed medians.}
    \label{fig:mixture-effect}
\end{figure}

A popular approach for assessing the overall mixture effect is to evaluate the exposure-risk surface at a range of exposure values. For instance, one might plot the fitted exposure-risk function while simultaneously setting all exposures to specific quantiles (see Figure \ref{fig:mixture-effect}). Using this strategy, the overall mixture effect suggests a decreasing risk of asthma-related ED visits with increased exposure levels. One challenge with this approach is that it is difficult to assess the contribution of each individual exposure to the overall mixture effect. A larger issue is that these exposure profiles are not particularly realistic. The observed pairwise proportions of exposures across all ZIP code days belonging to the same decile range from just 10\% (NO$_2$ and temperature, CO and temperature) to 23\% (NO$_2$ and CO). Meanwhile, just 0.06\% of all ZIP code days had all five exposures in the same decile. This observation underlines the need to evaluate the exposure-risk function in a more realistic manner.

\begin{figure}
    \centering
    \includegraphics[width=\textwidth]{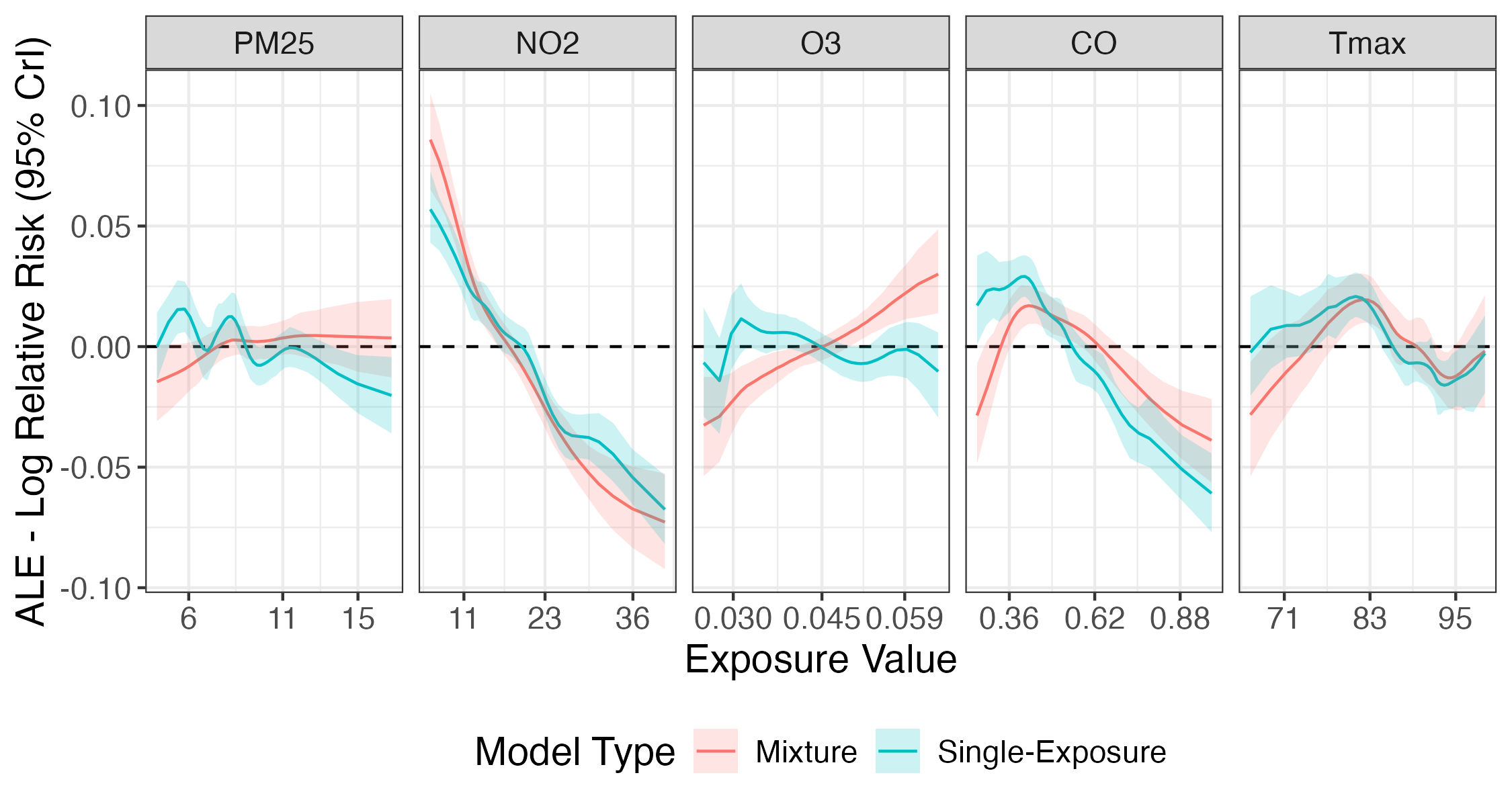}
    \caption{Main effect ALE for exposures individually (blue) and as part of a mixture (red) in the Atlanta application. Plots are centered so that zero on the y-axis represents an average risk level. ALEs are calculated using K = 40 quantile intervals for each exposure. Plots are trimmed so that only the central 95\% of each exposure is displayed.}
    \label{fig:asthma-main-effects}
\end{figure}

Alternatively, we can avoid extrapolation in our assessment of the mixture effect by referencing the ALE. Estimates of each exposure's ALE shift slightly in the mixture model compared to their single-exposure models (Figure \ref{fig:asthma-main-effects}). Most notably, the largely null effect of O$_3$ shifts to harmful in the mixture model. Additionally, in the mixture model, PM$_{2.5}$ has a borderline harmful main effect, NO$_2$ has a strong negative association with ED visits, and CO and temperature have some upside-down ``U-shaped" relationship with ED visits (Figure \ref{fig:asthma-main-effects}).

In the mixture model, the exposures may interact with one another as well. Here we focus on the potential joint effects of each chemical exposure with temperature, but the resulting fit also showed some interaction among the chemical exposures (see Appendix C of the Supplementary Material). In Figure \ref{fig:asthma-tmax-mod}, we note that the estimated ALE for each pollutant depends on temperature to some extent. For instance, the negative association between NO$_2$ and ED visits is more pronounced at lower temperatures. The positive association between O$_3$ and ED visits is more pronounced at lower temperatures, unless the O$_3$ concentration is very high. We also note that the estimated risk associated with PM$_{2.5}$ only appears to differ with temperature for lower PM$_{2.5}$ concentrations, while the reverse is true for CO.

\begin{figure}
    \centering
    \includegraphics[width=\textwidth]{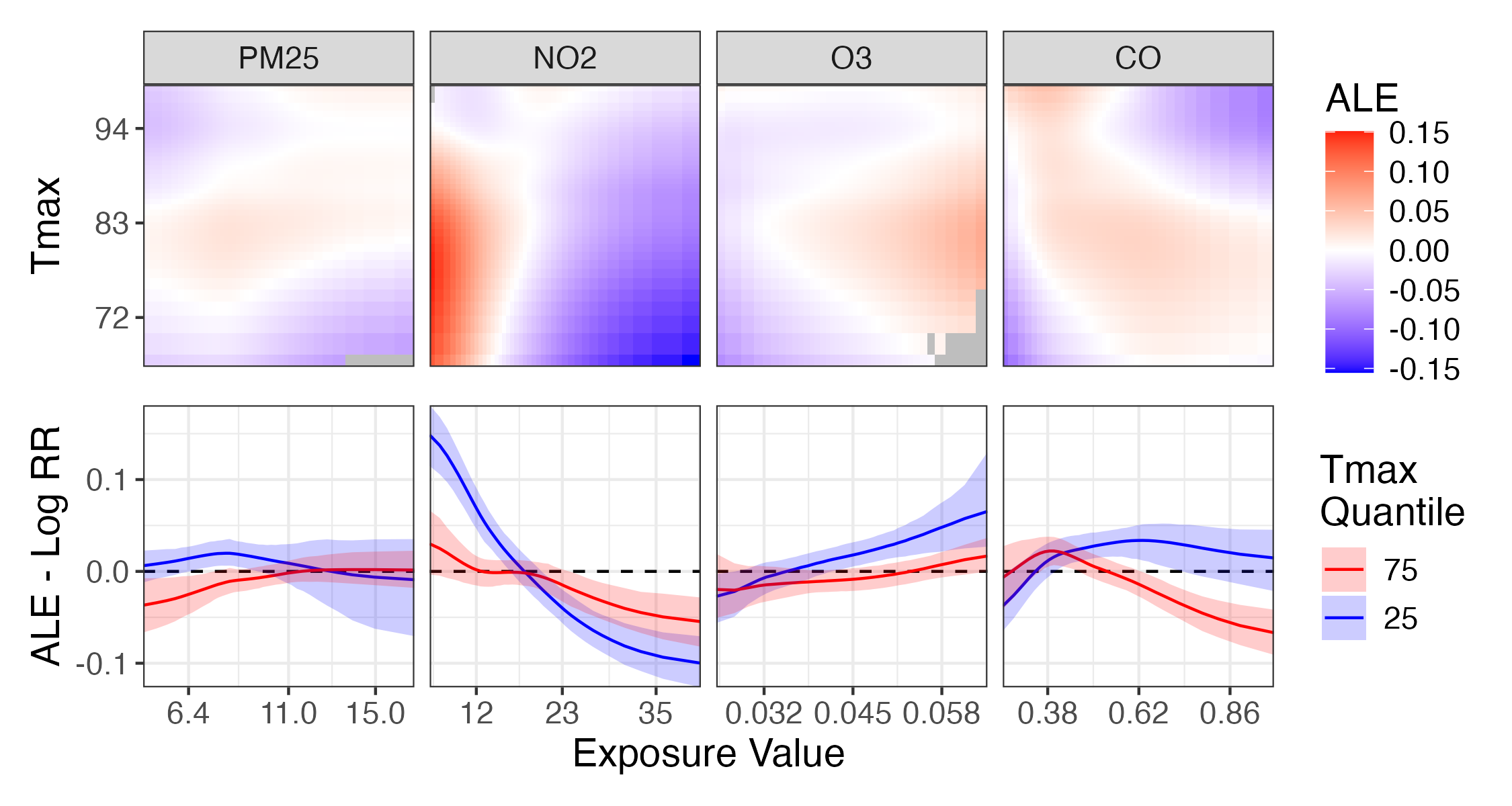}
    \caption{Pairwise second-order ALE for each chemical exposure with maximum temperature for the mixture model with $T = 25$, with the corresponding main effect ALEs added on. The top row includes the posterior mean ALE for all observed pairwise combinations, while the bottom row plots horizontal slices at the first and third quartiles of maximum temperature along with 95\% credible intervals. ALEs are calculated using K = 40 quantile intervals for each exposure. Plots are trimmed so that only the central 95\% of each exposure is displayed.}
    \label{fig:asthma-tmax-mod}
\end{figure}

\section{Discussion}
\label{sec:discuss}

In summary, we show through a simulation study and real data application that BART (specifically soft BART) can be used for estimating complex mixture exposure-risk surfaces in the context of count responses, such as visits to an emergency department. Additionally, we have demonstrated the utility of ALE for analyzing marginal and joint effects of 1-2 exposures in the presence of other exposures. Plots such as those included in Figures \ref{fig:asthma-main-effects} and \ref{fig:asthma-tmax-mod} are straightforward to interpret individually since the ALE estimation process only averages over plausible exposure profiles supported by the data.

Our application findings regarding asthma-related emergency department visits are also interesting. The exposure concentration required to achieve above average risk may depend on the temperature. We observed interaction between ozone and temperature - specifically a stronger ozone effect at cooler warm-season temperatures. While the analysis framework we have proposed is not causal in nature, we hypothesize that modification of chemical exposure effects by temperature could be related to individual-level behavior - e.g., people may be less likely to experience the effects of air pollution on very hot days where they are more inclined to stay indoors. We also found a strong negative association between NO$_2$ and ED visits, which stands in contrast to some findings regarding NO$_2$ and respiratory outcomes. Due to the dense tree canopy and high traffic emissions in Atlanta, NO$_2$ and volatile organic compounds are a precursor to ozone, and higher NO$_2$ levels may be reflective of warmer days with lower ozone pollution.

One current limitation of our methodology is the ability to formally detect lagged effects. Since in our application we are focused on short-term effects, using 3-day moving averages for the daily exposures is sufficient. In general it is difficult to simultaneously estimate non-linear, interaction, and lagged effects in mixture modeling. Regardless of the specific model selected, we find it important to consider exposures as a mixture rather than individually both when it comes to estimating the exposure-response surface and interpreting the resulting fit.

Our modeling approach is also computationally demanding. Drawing Pólya-gamma latent variables for every observation in the data augmentation step and updating the soft BART ensemble takes time. The model shared in the application took approximately 20 hours to fit and summarize. While this is a long time, Gaussian process based modeling approaches (e.g. BKMR) would be infeasible given the sample size of 220,000. A BART ensemble using traditional ``rigid" trees would also fit faster than soft BART, but may also require a greater number of trees to achieve comparable performance.

\section{Software}
\label{sec:software}

Software in the form of \texttt{R} code used to implement the proposed methods and produce the simulation results is available at \url{https://github.com/jacobenglert/softbart-mixtures-paper}.



\section*{Acknowledgments}
{\it Conflict of Interest}: None declared.

\section*{Funding}
The project is supported by the National Institutes of Health (P20ES036110, P30ES019776 and R01ES028346). The content is solely the responsibility of the authors and does not necessarily represent the official views of the NIH.  The data used to produce this publication were acquired from the Georgia Hospital Association. The contents of this publication, including data analysis, interpretation, conclusions derived, and the views expressed herein are solely those of the authors and do not represent the conclusions or official views of data sources listed above. Authorization to release this information does not imply endorsement of this study or its findings by any of these data sources. The data sources, their employees, officers, and agents make no representation, warranty, or guarantee as to the accuracy, completeness, currency, or suitability of the information provided here.

\bibliographystyle{biorefs}
\bibliography{references}

\end{document}


\title{Supplementary Material to \\ 
Modeling Joint Health Effects of Environmental Exposure Mixtures with Bayesian Additive Regression Trees}
\date{}

\author{JACOB R. ENGLERT$^\ast$ \\
\textit{Department of Biostatistics and Bioinformatics, Rollins School of Public Health, Emory University, 1518 Clifton Rd NE, Atlanta, GA 30322, USA} \\
jacob.englert@emory.edu \\
STEFANIE T. EBELT \\ 
\textit{Gangarosa Department of Environmental Health, Rollins School of Public Health, Emory University, 1518 Clifton Rd NE, Atlanta, GA 30322, USA}\\
HOWARD H. CHANG \\
\textit{Department of Biostatistics and Bioinformatics, Rollins School of Public Health, Emory University, 1518 Clifton Rd NE, Atlanta, GA 30322, USA}
}

\markboth%
{J. R. Englert and others}
{Modeling Joint Health Effects of Environmental Exposure Mixtures with BART}

\maketitle

\footnotetext{To whom correspondence should be addressed.}

\appendix

\section{MCMC Algorithm Details}

The Bayesian model described in Section 3 of the main text is fit with a Markov chain Monte-Carlo algorithm. The steps for this algorithm are outlined in this Appendix.

\subsection{Negative Binomial Representation}
The proposed model takes the following form:
\begin{gather}
    Y_{ij} \mid \theta_{ij} \sim \text{Poisson}(\theta_{ij}) \label{eq:outcome-regression} \\
    \theta_{ij} \sim \text{Gamma}(\xi, \exp{(\eta_{ij})}) \label{eq:theta-dist} \\
    \eta_{ij} = \log(\text{Pop}_{ij}) + \bX_{ij}^T \bbeta + f(\bZ_{ij}) + \nu_i \label{eq:log-predictor}.
\end{gather}

As mentioned in the manuscript, this hierarchical model can be shown to have a marginal negative binomial distribution with density given by

\begin{equation}
\label{eq:nb-density}
    p (y_{ij} \mid \xi, \eta_{ij}) = \frac{\Gamma(y_{ij} + \xi)}{\Gamma(y_{ij} + 1)\Gamma(\xi)} (1-p_{ij})^\xi p_{ij}^{y_{ij}} \propto (1-p_{ij})^\xi p_{ij}^{y_{ij}}
\end{equation}
where $p_{ij} = \frac{\exp{(\eta_{ij})}}{1 + \exp{(\eta_{ij})}}$. The mean and variance may be obtained in a straightforward fashion from the original formulation:
\begin{equation}
    \mathbb{E}(Y_{ij}) = \mathbb{E}\left\{\mathbb{E}(Y_{ij} \mid \theta_{ij}) \right\} = \mathbb{E} (\theta_{ij}) = \xi \exp{(\eta_{ij})}
\end{equation}
\begin{equation}
\begin{split}
    \Var(Y_{ij}) & = \Var \left\{\mathbb{E}(Y_{ij} \mid \theta_{ij})\right\} + \mathbb{E}\left\{\Var(Y_{ij} \mid \theta_{ij})\right\} \\
    & = \Var (\theta_{ij}) + \mathbb{E}(\theta_{ij}) \\
    & = \xi \exp{(2 \eta_{ij})} + \xi \exp{(\eta_{ij})} \\
    & = \xi \exp{(\eta_{ij})} \left\{ 1 + \exp{(\eta_{ij})} \right\}
\end{split}
\end{equation}

Since $\Var(Y_{ij}) > \mathbb{E}(Y_{ij})$, it is clear that $\xi$ is exclusively modeling \emph{overdispersion}. The marginal likelihood of all of the observed data can be written as in \eqref{eq:nb-total-likelihood}.

\begin{equation}
\label{eq:nb-total-likelihood}
    p (\biy \mid \xi, \boldeta) \propto \prod_{i=1}^I \prod_{j=1}^J (1-p_{ij})^\xi p_{ij}^{y_{ij}} = \prod_{i=1}^I \prod_{j=1}^J \frac{\left\{\exp(\eta_{ij})\right\}^{y_{ij}}}{\left\{1 + \exp(\eta_{ij})\right\}^{y_{ij} + \xi}}.
\end{equation}

\subsection{Pólya-Gamma Data Augmentation}
BART (and by extension, soft BART) typically requires conditional conjugacy between the outcome distribution and prior distribution on the individual leaf node parameters to facilitate the sampling of tree structures. Since the negative binomial likelihood in Equation \eqref{eq:nb-density} does not admit a conditionally conjugate prior, we adopt the framework proposed by \citet{pillow_fully_2012}. Specifically, we augment the outcome $Y_{ij}$ with latent weights $\omega_{ij}$ sampled from a Pólya-gamma (PG) distribution \citep{polson_bayesian_2013}.

If $\omega \sim \text{PG}(b, 0)$, then for any choice of $a$, we have the following result:

\begin{equation}
    \frac{(e^\eta)^a}{(1 + e^\eta)^b} = 2^{-b}e^{\kappa\eta} \int_0^\infty e^{-\omega \eta^2 / 2} p(\omega \mid b, 0) d \omega
\end{equation}

where $\kappa = a - b/2$. Substituting $\eqref{eq:nb-total-likelihood}$ into the LHS of this result gives

\begin{equation}
    p (\biy \mid \xi, \boldeta) \propto \prod_{i=1}^I \prod_{j=1}^{J} \exp{(\kappa_{ij} \eta_{ij})} \int_0^\infty \exp{(-\omega_{ij}\eta_{ij}^2/2)} p(\omega_{ij} \mid y_{ij} + \xi, 0) d\omega_{ij}
\end{equation}

where $\kappa_{ij} = \frac{y_{ij} - \xi}{2}$. If we condition on $\omega_{ij}$, the expectation (integral) in the above expression can be ignored and we have
\begin{equation}
\begin{split}
    p(\biy \mid \xi, \boldeta, \bomega) & \propto \prod_{i=1}^I \prod_{j=1}^{J} \exp{(\kappa_{ij} \eta_{ij})} \times \exp{(-\omega_{ij} \eta_{ij}^2 / 2)} \\
    & \propto \prod_{i=1}^I \prod_{j=1}^{J} \exp{\left\{ -\frac{\omega_{ij}}{2}\left(\frac{y_{ij} - \xi}{2\omega_{ij}} - \eta_{ij}\right)^2 \right\}},
\end{split}
\end{equation}
where $\bomega = (\omega_{11}, \ldots, \omega_{1J}, \ldots, \omega_{I1}, \ldots, \omega_{IJ})^T$. Thus, if we independently draw $\omega_{ij} \sim \text{PG}(Y_{ij} + \xi, \eta_{ij})$, then the latent outcome $y^*_{ij} = \frac{y_{ij} - \xi}{2\omega_{ij}}$ is normally distributed with mean $\eta_{ij}$ and variance $1 / \omega_{ij}$. It follows that the vector of latent outcomes $\biy^* \sim \text{MVN}(\boldeta, \bOmega^{-1})$. This leads to a convenient Gibbs sampler based on the Bayesian backfitting approach of \citet{hastie_bayesian_2000} for updating $\bbeta$, $\left\{\cM_t \right\}_{t=1}^T$, and $\bnu$.

\subsection{Updating $\bbeta$}

Let $r^\beta_{ij} = y^*_{ij} - \log \left(\text{Pop}_{ij}\right) - \sum_{t=1}^T g(\bZ_{ij} \mid \cT_t, \cM_t) - \nu_i$, and define $\bir^\beta = (r^\beta_{11}, \ldots, r^\beta_{IJ})^T$. Then, assuming $\pi (\bbeta) = \text{MVN}(\mathbf{b}, \bSigma_\beta)$, we derive the posterior distribution for $\bbeta$ as:
\begin{equation}
\begin{split}
    \pi (\bbeta \mid \biy, \boldeta, \bomega, \xi) & \propto \pi(\bbeta) \times p (\bir^\beta \mid \bbeta, \bOmega) \\
    & \propto \exp{\left\{-\frac{1}{2} \left(\bbeta - \mathbf{b} \right\}^T \bSigma_\beta^{-1} \left(\bbeta - \mathbf{b} \right) \right\}} \times \exp{\left\{-\frac{1}{2} \left(\bX\bbeta - \bir^\beta \right)^T \bOmega \left(\bX\bbeta - \bir^\beta \right) \right\}} \\
    & \propto \exp{\left[ -\frac{1}{2} \left\{(\bbeta - \mathbf{b})^T \bSigma_\beta^{-1} (\bbeta - \mathbf{b} ) + \left(\bX\bbeta - \bir^\beta \right)^T \bOmega \left(\bX\bbeta - \bir^\beta \right) \right\} \right]} \\
    & \propto \text{Normal}\left(\bmu_\beta^*, \bSigma_\beta^* \right)
\end{split}
\end{equation}
where $\bSigma_\beta^* = \left(\bSigma_\beta^{-1} + \bX^T \bOmega \bX \right)^{-1}$ and $\bmu_\beta^* = \bSigma_\beta^* \left(\bSigma_\beta^{-1} \mathbf{b} + \bX^T \bOmega \bir^\beta \right)$.

\subsection{Updating $\bnu$}
Let $r^\nu_{ij} = y^*_{ij} - \log \left(\text{Pop}_{ij}\right) - \bX_{ij}^T\bbeta - \sum_{t=1}^T g(\bZ_{ij} \mid \cT_t, \cM_t)$, and define $\bir^\nu = (r^\nu_{11}, \ldots, r^\nu_{IJ})^T$. Recall that we assume a proper CAR prior for $\bnu$, i.e. $\bnu \sim \text{MVN}(\bzero, \bSigma_\nu)$ where $\bSigma_\nu = \tau^2 \left(\bD - \rho\bW)^{-1} \right)$. $\bW$ is the $I \times I$ first-order adjacency matrix, and $\bD$ is the $I \times I$ diagonal matrix containing the number of neighbors for each region. Let $\bX_\nu$ be the $IJ \times I$ design matrix such that $\bX_\nu \bnu$ is the $IJ \times 1$ vector of spatial random effects allocated to each observation in the dataset. We derive the posterior distribution for $\bnu$ as:
\begin{equation}
\begin{split}
    \pi (\bnu \mid \biy, \boldeta, \bomega, \xi) & \propto \pi(\bnu) \times p (\bir^\nu \mid \bnu, \bOmega) \\
    & \propto \exp{\left(-\frac{1}{2} \bnu^T \bSigma_\nu^{-1} \bnu \right)} \times \exp{\left\{-\frac{1}{2} \left(\bX_\nu\bnu - \bir^\nu \right)^T \bOmega \left(\bX_\nu\bnu - \bir^\nu \right) \right\}} \\
    & \propto \exp{\left[ -\frac{1}{2} \left\{ \bnu^T \bSigma_\nu^{-1} \bnu + \left(\bX_\nu\bnu - \bir^\nu \right)^T \bOmega \left(\bX_\nu\bnu - \bir^\nu \right) \right\} \right]} \\
    & \propto \text{Normal}\left(\bmu_\nu^*, \bSigma_\nu^* \right)
\end{split}
\end{equation}
where $\bSigma_\nu^* = \left(\bSigma_\nu^{-1} + \bX_\nu^T \bOmega \bX_\nu \right)^{-1}$ and $\bmu_\nu^* = \bSigma_\nu^* \bX_\nu^T \bOmega \bir^\nu$.

\subsection{Updating $\tau^2$}
Assuming $\pi (\tau^2) = \text{Inverse-Gamma}\left(\alpha_\tau, \beta_\tau \right)$, the posterior distribution of $\tau^2$ is derived as:
\begin{equation}
\begin{split}
\pi (\tau^2 \mid \bnu) & \propto \pi(\tau^2) \times p (\bnu \mid \tau^2) \\
& \propto \frac{{\beta_\tau}^{\alpha_\tau}}{\Gamma(\alpha_\tau)} (\tau^2)^{-\alpha_\tau - 1} \exp{\left(-\frac{\beta_\tau}{\tau^2} \right)} \times \frac{1}{\sqrt{\det \left(\bSigma_\nu \right)}} \times \exp{\left(-\frac{1}{2} \bnu^T \bSigma_\nu^{-1} \bnu \right)} 
\end{split}
\end{equation}

If we assume $\bSigma_\nu = \tau^2(\bD - \rho \bW)^{-1}$, as is the case when using a proper CAR prior for $\bnu$, then we have:
\begin{equation}
\begin{split}
\pi(\tau^2 \mid \bnu ) & \propto \left(\tau^2 \right)^{-\left(\alpha_\tau + \frac{I}{2} \right) - 1} \exp{\left[-\frac{1}{\tau^2} \left\{\beta_\tau + \frac{\bnu^T(\bD - \rho \bW)\bnu}{2} \right\} \right]} \\
& \propto \text{Inverse-Gamma}\left( \alpha_\tau + \frac{I}{2}, \quad \beta_\tau + \frac{\bnu^T(\bD - \rho \bW)\bnu}{2} \right)
\end{split}
\end{equation}

where the above uses the result that for constant $c$ and square matrix $\bA_{N \times N}$, we have $\det{(c \bA)} = c^{N} \det{(\bA)}$.

\subsection{Updating $\rho$}
For updating $\rho$, we assign a discrete uniform prior over 1,000 values between 0 and 1 (i.e., $\rho \sim \text{Unif}\left\{\frac{0}{999}, \ldots, \frac{999}{999} \right\}$). Since this prior assigns equal probability to all values in its support, the full conditional distribution for $\rho$ is proportional to the density of the random effects \eqref{eq:rho-density}.
\begin{equation}
\label{eq:rho-density}
    \pi \left( \rho \mid \bnu, \tau^2 \right) \propto \pi \left( \rho \right) \times p \left( \bnu \mid \rho, \tau^2 \right) \propto p \left(\bnu \mid \rho, \tau^2 \right)
\end{equation}

Recall that we assume a proper CAR prior for $\bnu$. In other words, $\bnu \sim \text{MVN}(\bzero, \bSigma_\nu)$, where $\bSigma_\nu = \tau^2(\bD - \rho \bW)^{-1}$. Thus, the log density of $p (\bnu \mid \tau^2, \rho)$ is given by
\begin{equation}
\label{eq:rho-update}
\begin{split}
    \log \pi \left(\bnu \mid \tau^2, \rho \right) & \propto \frac{1}{2} \log | \tau^2(\bD - \rho \bW )^{-1} | - \frac{1}{2\tau^2}\bnu^T(\bD - \rho\bW)\bnu \\
    & \propto \frac{1}{2}\log | \bD - \rho \bW| + \frac{\rho}{2\tau^2}\bnu^T \bW\bnu \\
    & = \frac{1}{2}\log | \bD(\mathbf{I} - \rho \bD^{-1}\bW)| + \frac{\rho}{2\tau^2}\bnu^T \bW\bnu \\
    & \propto \frac{1}{2}\log | \mathbf{I} - \rho \bD^{-1}\bW | + \frac{\rho}{2\tau^2}\bnu^T \bW\bnu \\
    & = \frac{1}{2}\sum_{i=1}^I \log(1 - \rho \lambda_i) + \frac{\rho}{2\tau^2}\bnu^T \bW\bnu
\end{split}
\end{equation}

where $\lambda_i$ is the $i^{th}$ eigenvalue of $\bD^{-1}\bW$. The first term in the final expression may be calculated ahead of time for all candidate values of $\rho$ so that only the second term needs to be updated during the MCMC. New values of $\rho$ can be sampled from the discrete distribution with probabilities proportional to the computed values of \eqref{eq:rho-update} for all 1,000 prior values of $\rho$. Alternatively, an intrinsic CAR prior may be used, where $\rho$ if fixed to 1 throughout the MCMC algorithm. In many applications the posterior distribution of $\rho$ will be concentrated around 1, but using the approach detailed here will provide more flexibility when that is not the case.

\subsection{Updating BART Parameters}

In BART, new tree structures are sampled from their marginal distribution and updated using an MH step by first integrating out the leaf node parameters. A detailed explanation of how this approach works with weights, such as those generated by our data augmentation scheme, is outlined in \citet{bleich_bayesian_2014}. The tree prior used for soft BART is different, but the general concept is the same. One of the benefits of BART is that default priors tend to work well in a variety of circumstances. We use default priors for updating $\left\{\cT_t, \cM_t \right\}_{t=1}^T$ as detailed in \citet{linero_bayesian_2018-1}.

\subsection{Updating $\xi$}
The final update in the MCMC algorithm is for the (over)dispersion parameter $\xi$. We perform this update using the conjugate sampling routine described in \cite{zhou_lognormal_2012}.  This technique relies on expressing the $Y_{ij} \sim \text{NB}(\xi, p_{ij})$ marginal distribution as a compound Poisson distribution \eqref{eq:compound-poisson}.
\begin{gather}
    y_{ij} = \sum_{m=1}^{L_{ij}} l_m \nonumber \\
    L_{ij} \sim \text{Poisson}(-\xi \ln(1-p_{ij})) \label{eq:compound-poisson}
\end{gather}

Given $y_{ij}$ and $\xi$, $L_{ij}$ follows a Chinese Restaurant Table (CRT) distribution whose samples can be generated as $L_{ij} = \sum_{m=1}^{y_{ij}} l_m$ where $l_m \sim \text{Bernoulli}\left(\frac{\xi}{\xi + m - 1} \right)$. If we assign a $\text{Gamma}(\alpha_\xi, \beta_\xi)$ prior for $\xi$, the full conditional distribution for $\xi$ is derived as:
\begin{equation}
\label{eq:xi-update}
\begin{split}
    \pi (\xi \mid p ) & \propto \pi(\xi) \prod_{i=1}^I \prod_{j=1}^J p (L_{ij} \mid \xi, p_{ij}) \\
    & = \frac{\beta_\xi^{\alpha_\xi}}{\Gamma(\alpha_\xi)}\xi^{\alpha_\xi - 1} \exp{(-\xi \beta_\xi)} \times \prod_{i=1}^I \prod_{j=1}^J \frac{\left\{-\xi \ln(1-p_{ij})\right\}^{L_{ij}} \exp{\left\{\xi \ln(1-p_{ij})\right\}}}{L_{ij}!} \\
    & \propto \xi^{\alpha_\xi - 1} \exp{(-\xi \beta_\xi)} \times \xi^{\sum \sum L_{ij}} \exp{ \left\{ \xi \sum_{i=1}^I \sum_{j=1}^J \ln(1-p_{ij}) \right\}} \\
    & = \xi^{\alpha_\xi + \sum \sum L_{ij} - 1}\exp{\left[-\xi \left\{\beta_\xi - \sum_{i=1}^I \sum_{j=1}^J \ln(1-p_{ij}) \right\} \right]} \\
    & \propto \text{Gamma}\left(\alpha_\xi + \sum_{i=1}^I \sum_{j=1}^J L_{ij}, \beta_\xi - \sum_{i=1}^I \sum_{j=1}^J \ln(1-p_{ij}) \right).
\end{split}
\end{equation}

See \cite{zhou_lognormal_2012} for further details regarding this procedure.

\section{Additional Simulation Results}
\renewcommand{\thefigure}{B\arabic{figure}}
\setcounter{figure}{0}
\renewcommand{\thetable}{B\arabic{table}}
\setcounter{table}{0}

\begin{figure}[H]
    \centering
    \includegraphics[width=\linewidth]{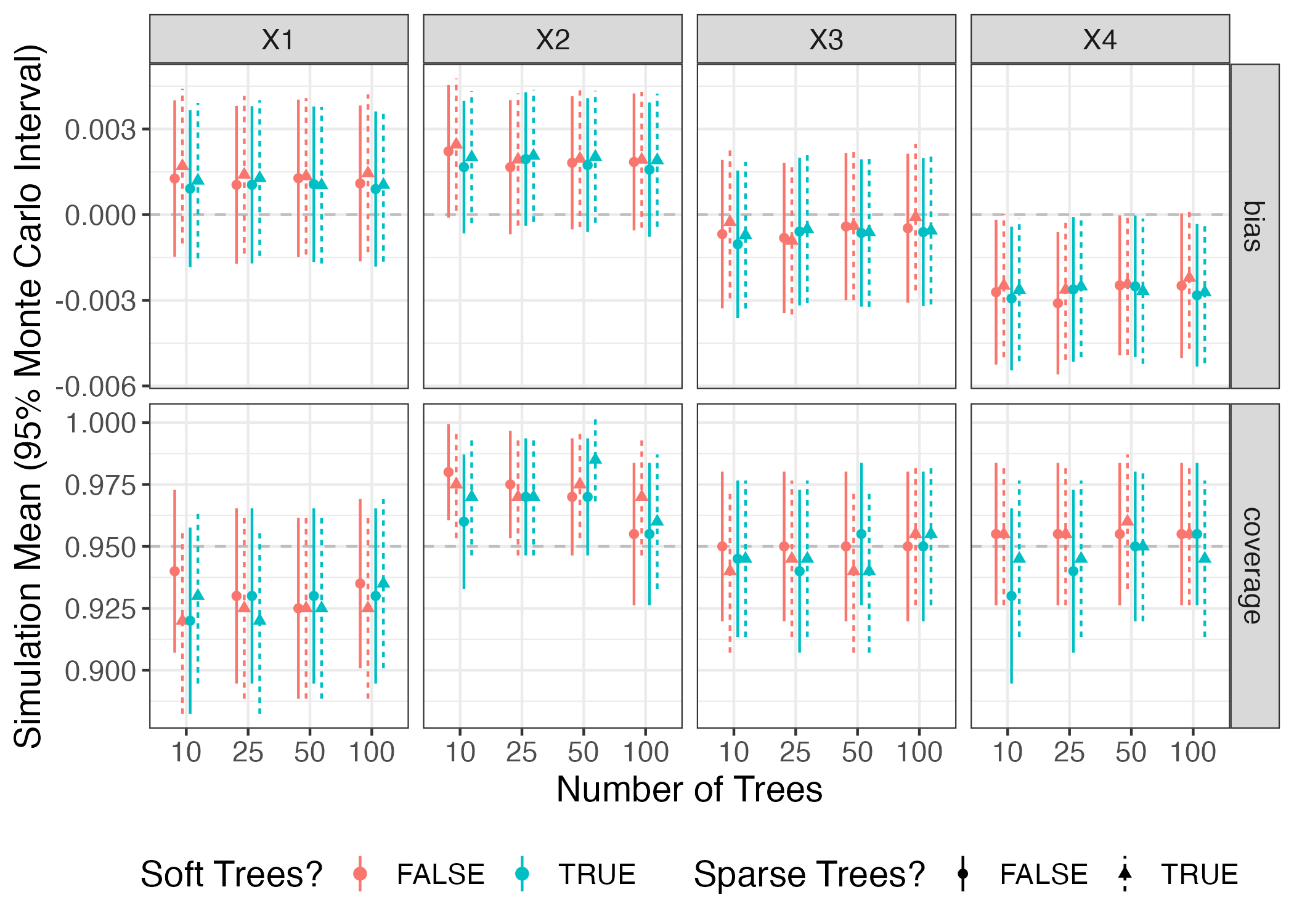}
    \caption{Bias and 95\% credible interval coverage for the four linearly modeled parameters in the simulation study. Estimates are plotted as the simulation mean +/- $1.96 \times$ the simulation Monte Carlo standard error.}
    \label{fig:sim-study-beta-stats}
\end{figure}

\begin{sidewaystable}

\caption{\label{tab:sim-global-stats}Bias and 95\% CrI coverage for global parameters in simulation study.}
\centering
\begin{tabular}[t]{ccccccccc}
\toprule
\multicolumn{3}{c}{ } & \multicolumn{2}{c}{$\rho$} & \multicolumn{2}{c}{$\tau^2$} & \multicolumn{2}{c}{$\xi$} \\
\cmidrule(l{3pt}r{3pt}){4-5} \cmidrule(l{3pt}r{3pt}){6-7} \cmidrule(l{3pt}r{3pt}){8-9}
$T$\footnotemark[1] & Soft\footnotemark[2] & Sparse\footnotemark[3] & Bias (MCSE) & Coverage (MCSE) & Bias (MCSE) & Coverage (MCSE) & Bias (MCSE) & Coverage (MCSE)\\
\midrule
10 &  &  & -0.099 (0.008) & 0.905 (0.021) & 0.007 (0.003) & 0.965 (0.013) & -0.020 (0.000) & 0.205 (0.029)\\
10 &  & \checkmark & -0.098 (0.008) & 0.910 (0.020) & 0.007 (0.003) & 0.970 (0.012) & -0.022 (0.000) & 0.190 (0.028)\\
10 & \checkmark &  & -0.098 (0.008) & 0.910 (0.020) & 0.007 (0.003) & 0.970 (0.012) & -0.004 (0.000) & 0.930 (0.018)\\
10 & \checkmark & \checkmark & -0.099 (0.008) & 0.905 (0.021) & 0.007 (0.003) & 0.965 (0.013) & -0.004 (0.000) & 0.960 (0.014)\\
\addlinespace
25 &  &  & -0.099 (0.008) & 0.910 (0.020) & 0.007 (0.003) & 0.965 (0.013) & -0.009 (0.000) & 0.775 (0.030)\\
25 &  & \checkmark & -0.098 (0.008) & 0.905 (0.021) & 0.007 (0.003) & 0.975 (0.011) & -0.009 (0.000) & 0.775 (0.030)\\
25 & \checkmark &  & -0.098 (0.008) & 0.900 (0.021) & 0.007 (0.003) & 0.975 (0.011) & -0.003 (0.000) & 0.945 (0.016)\\
25 & \checkmark & \checkmark & -0.099 (0.008) & 0.910 (0.020) & 0.007 (0.003) & 0.980 (0.010) & -0.003 (0.000) & 0.955 (0.015)\\
\addlinespace
50 &  &  & -0.098 (0.008) & 0.915 (0.020) & 0.007 (0.003) & 0.970 (0.012) & -0.005 (0.000) & 0.930 (0.018)\\
50 &  & \checkmark & -0.098 (0.008) & 0.910 (0.020) & 0.006 (0.003) & 0.970 (0.012) & -0.005 (0.000) & 0.905 (0.021)\\
50 & \checkmark &  & -0.099 (0.008) & 0.900 (0.021) & 0.007 (0.003) & 0.970 (0.012) & -0.003 (0.000) & 0.950 (0.015)\\
50 & \checkmark & \checkmark & -0.098 (0.008) & 0.905 (0.021) & 0.006 (0.003) & 0.965 (0.013) & -0.003 (0.000) & 0.950 (0.015)\\
\addlinespace
100 &  &  & -0.099 (0.008) & 0.915 (0.020) & 0.007 (0.003) & 0.970 (0.012) & -0.003 (0.000) & 0.960 (0.014)\\
100 &  & \checkmark & -0.099 (0.008) & 0.895 (0.022) & 0.006 (0.003) & 0.975 (0.011) & -0.003 (0.000) & 0.960 (0.014)\\
100 & \checkmark &  & -0.098 (0.008) & 0.895 (0.022) & 0.006 (0.003) & 0.960 (0.014) & -0.003 (0.000) & 0.975 (0.011)\\
100 & \checkmark & \checkmark & -0.098 (0.008) & 0.920 (0.019) & 0.007 (0.003) & 0.965 (0.013) & -0.003 (0.000) & 0.970 (0.012)\\
\bottomrule
\multicolumn{9}{l}{\rule{0pt}{1em}CrI: Bayesian posterior credible interval.}\\
\multicolumn{9}{l}{\rule{0pt}{1em}MCSE: Monte Carlo Standard Error.}\\
\multicolumn{9}{l}{\rule{0pt}{1em}\textsuperscript{*} Number of trees.}\\
\multicolumn{9}{l}{\rule{0pt}{1em}\textsuperscript{\dag} Soft BART used \citep{linero_bayesian_2018-1}.}\\
\multicolumn{9}{l}{\rule{0pt}{1em}\textsuperscript{\ddag} Sparse branching process used \citep{linero_bayesian_2018}.}\\
\end{tabular}
\end{sidewaystable}


\begin{figure}[H]
    \centering
    \includegraphics[width=\linewidth]{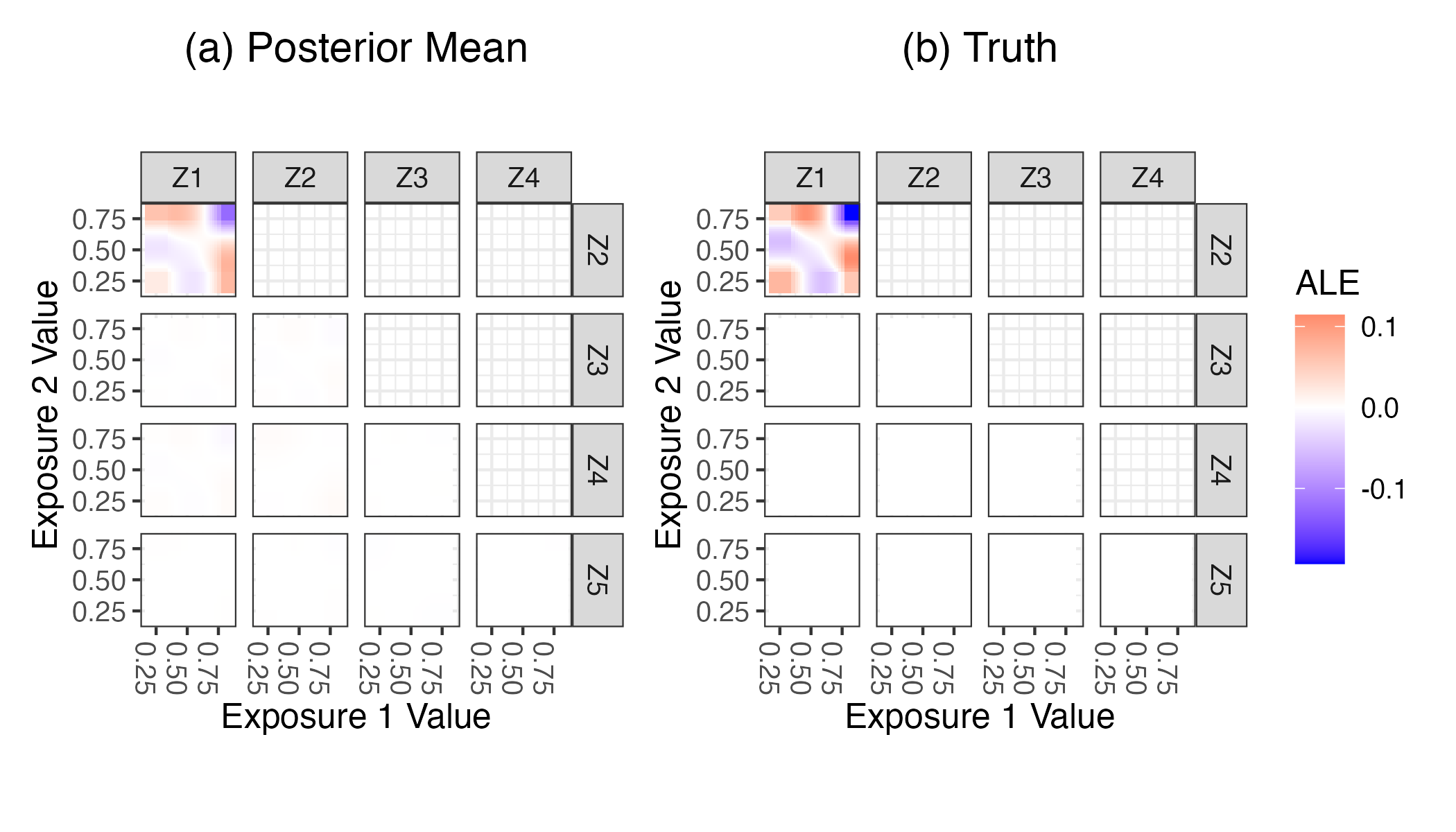}
    \caption{ALE interaction (second-order only) effects from the simulation study with $T = 25$, soft, sparse trees. Plot (a) displays the simulation average posterior mean at each grid cell, while (b) displays the truth calculated per the data generating process based on \cite{friedman_multivariate_1991}. ALEs are calculated using $K=40$ quantile intervals for each exposures. Plots are trimmed so that only the central 95\% of each exposure is displayed.}
    \label{fig:sim-study-interaction-effects}
\end{figure}

\begin{figure}[H]
    \centering
    \includegraphics[width=\linewidth]{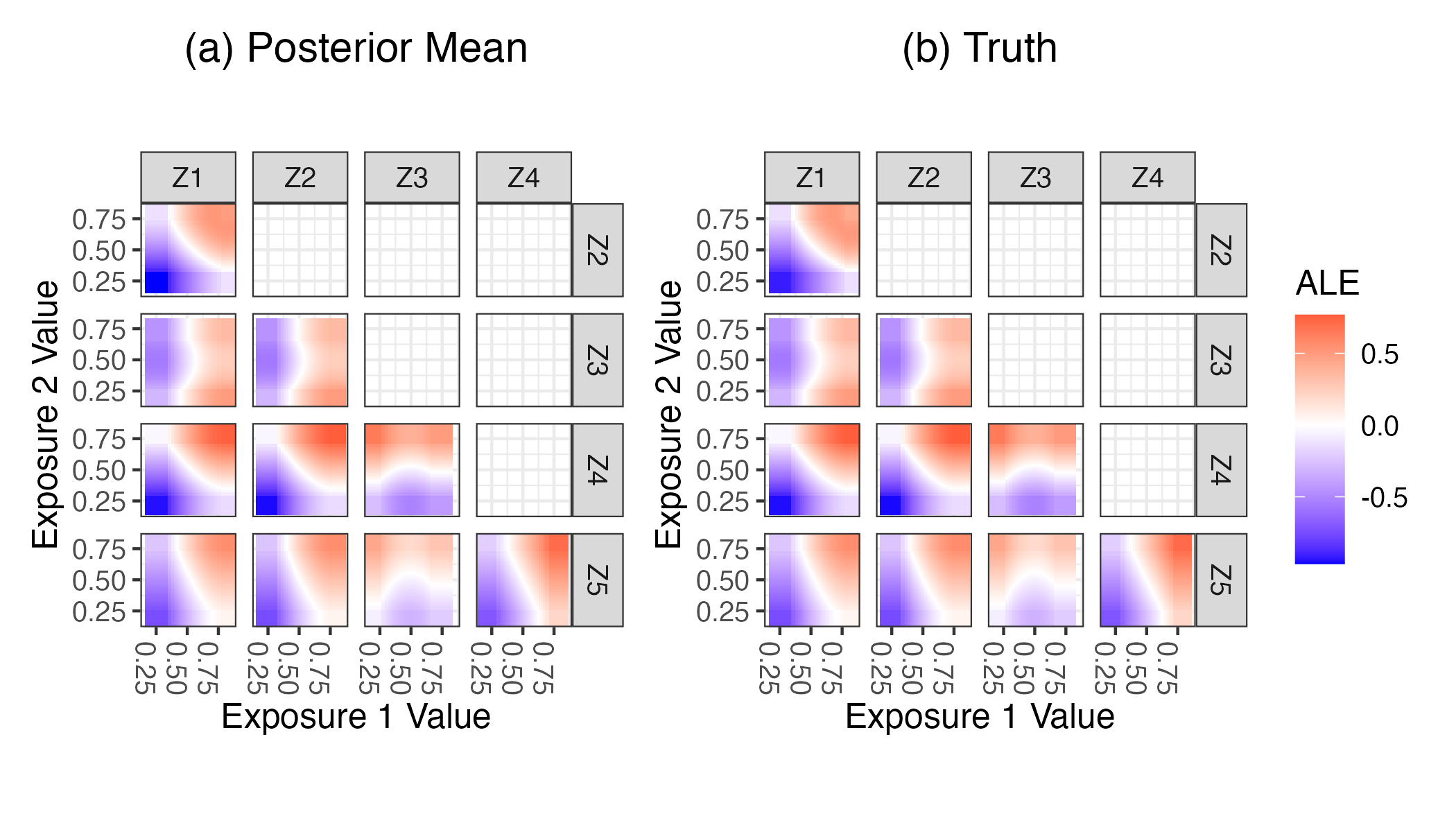}
    \caption{ALE pairwise joint effects (main effects + second-order effects) from the simulation study with $T = 25$, soft, sparse trees. Plot (a) displays the simulation average posterior mean at each grid cell, while (b) displays the truth calculated per the data generating process based on \cite{friedman_multivariate_1991}. ALEs are calculated using $K=40$ quantile intervals for each exposures. Plots are trimmed so that only the central 95\% of each exposure is displayed.}
    \label{fig:sim-study-joint-effects}
\end{figure}

\newpage

\section{Additional Application Results}
\renewcommand{\thefigure}{C\arabic{figure}}
\setcounter{figure}{0}
\renewcommand{\thetable}{C\arabic{table}}
\setcounter{table}{0}

\begin{figure}[H]
    \centering
    \includegraphics[width=0.9\linewidth]{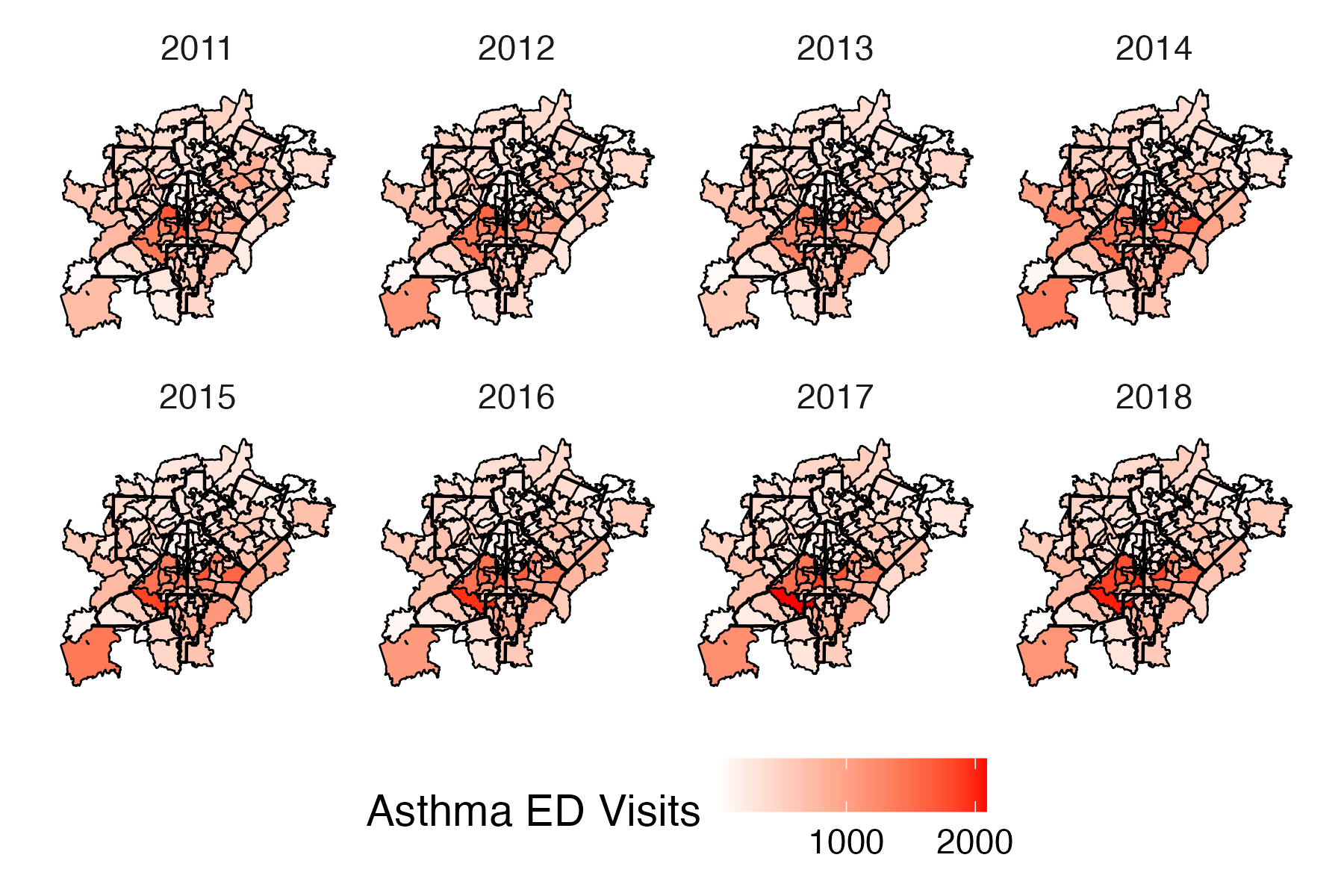}
    \caption{Annual ZIP code-level counts of asthma-related emergency department visits.}
    \label{fig:annual-asthma-maps}
\end{figure}

\begin{figure}[H]
    \centering
    \includegraphics[width=0.9\linewidth]{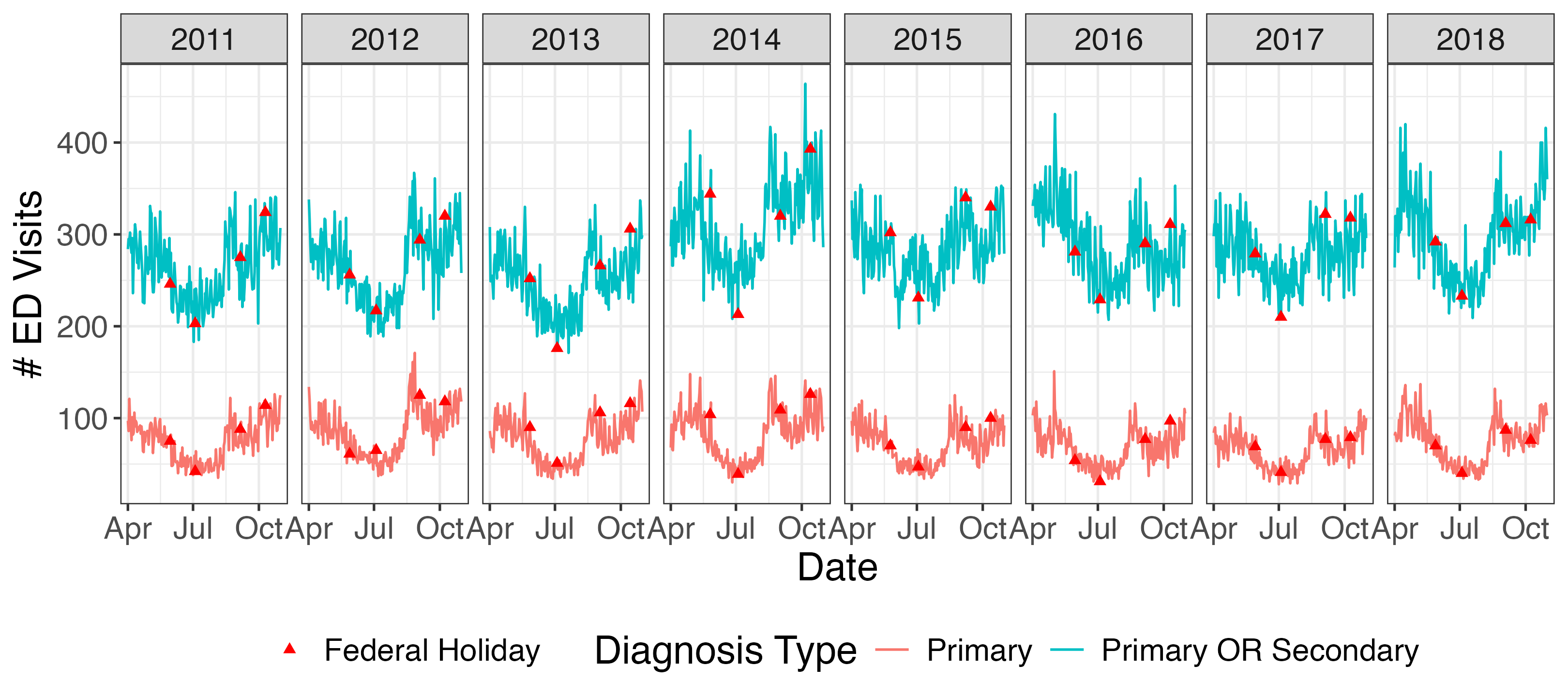}
    \caption{Daily counts of asthma-related emergency department (ED) visits, aggregated over ZIP code. Counts of ED visits with asthma as the primary diagnosis are shown in red, while visits with any asthma diagnosis are shown in blue.}
    \label{fig:asthma-time-series}
\end{figure}

\begin{figure}[H]
    \centering
    \includegraphics[width=\linewidth]{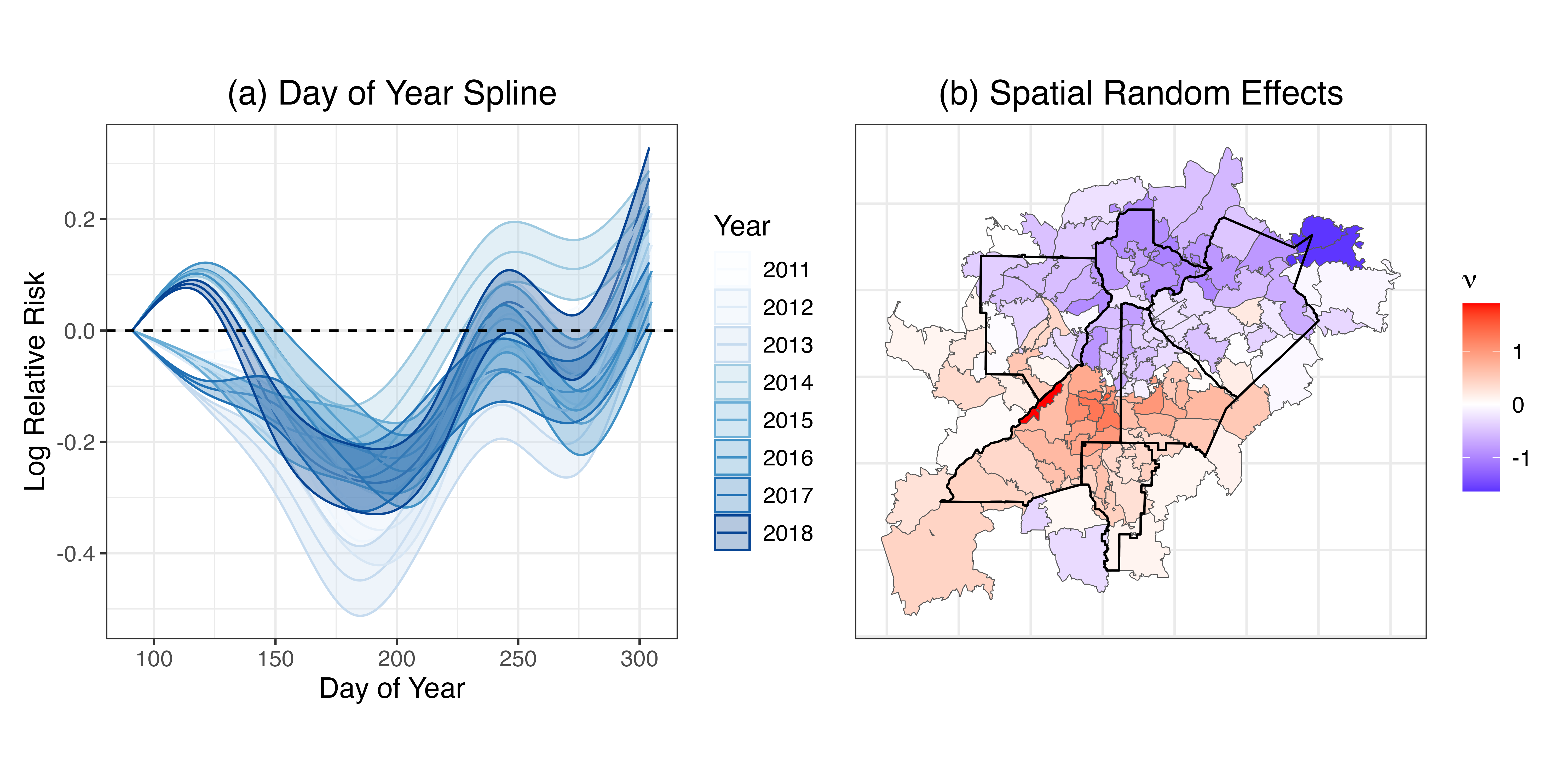}
    \caption{Estimated day-of-year spline (a) and posterior mean ZIP code-level random effects (b) obtained from the 25-tree mixture model for the Atlanta ED visit data. Both are presented on the log scale.}
    \label{fig:doy-spline-and-random-effects}
\end{figure}

\begin{figure}[H]
    \centering
    \includegraphics[width=\linewidth]{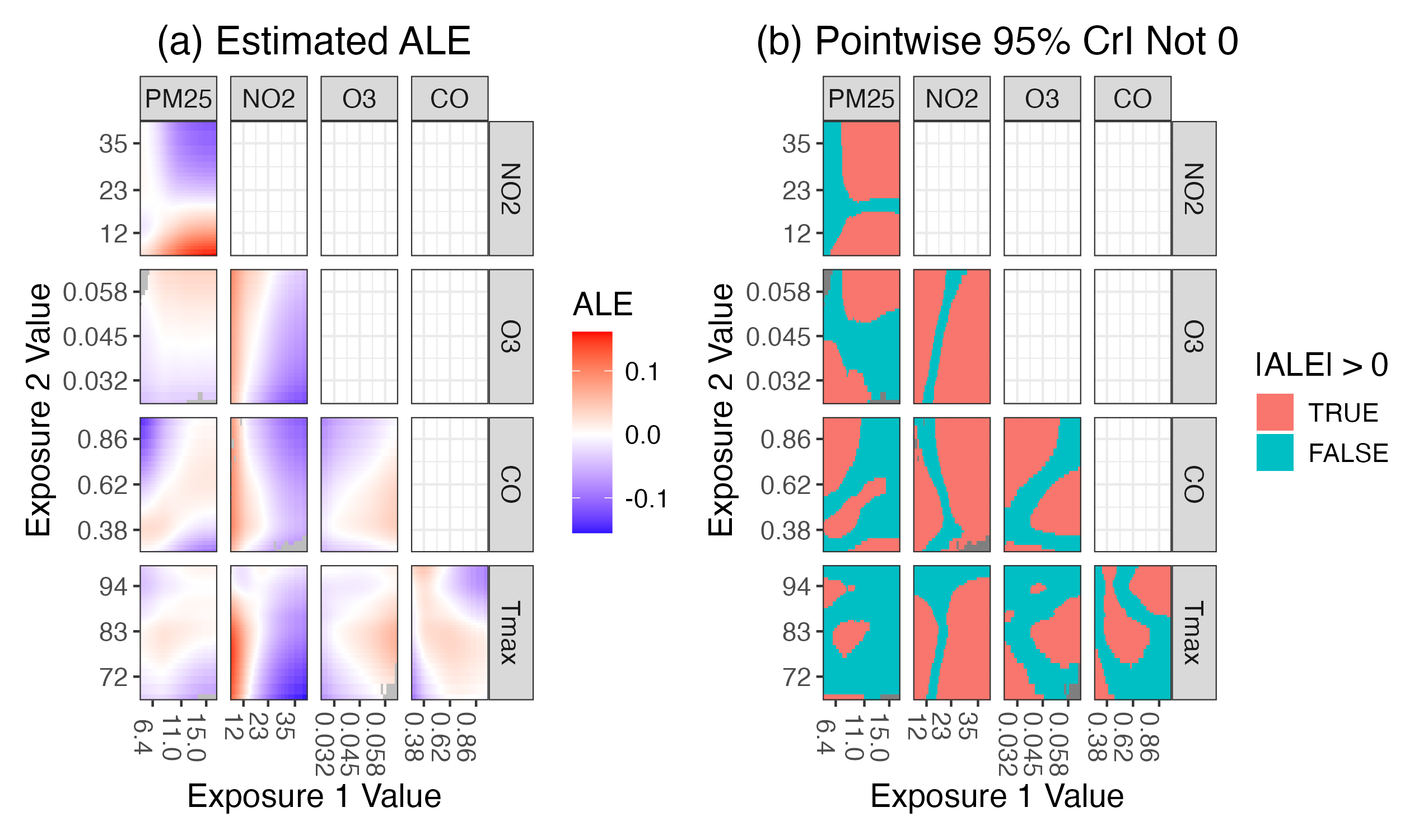}
    \caption{All pairwise second-order ALE for the mixture model with $T = 25$, with the corresponding main effect ALEs added on. The posterior means for each observed pairwise combination of exposures are plotted in (a), while (b) indicates whether or not the corresponding pointwise 95\% credible intervals contain 0. ALEs are calculated using K = 40 quantile intervals for each exposure. Plots are trimmed so that only the central 95\% of each exposure is displayed.}
    \label{fig:all-asthma-second-order-ale}
\end{figure}

\section{Implementation Details}

\texttt{R} code and packages used to implement the proposed methods and produce the simulation results is available at \url{https://github.com/jacobenglert/softbart-mixtures-paper}. The sampling of BART-related parameters is carried out using the drop-in C++ (Rcpp) module available in the \texttt{SoftBart} package. Due to prolonged runtimes, all analyses (simulation and real data) were originally run on the high performance computing cluster at the Rollins School of Public Health at Emory University.

\bibliographystyle{biorefs}
\bibliography{references}